\documentclass[12pt]{article}
\usepackage{fullpage,setspace}
\usepackage{amssymb, amsthm, amsmath}
\usepackage{graphicx}
\usepackage[authoryear]{natbib}
\usepackage{bm}
\usepackage{times}
\usepackage{setspace}
\usepackage{wrapfig}
\usepackage{epstopdf}
\usepackage{subcaption}
\usepackage{caption}
\usepackage{multirow}
\usepackage{verbatim}
\usepackage{url}

\newcommand{\btheta}{ \mbox{\boldmath $\theta$}}

\newcommand{\bbeta}{ \mbox{\boldmath $\beta$}}

\newcommand{\bpsi}{ \mbox{\boldmath $\psi$}}
\newcommand{\bphi}{ \mbox{\boldmath $\phi$}}

\newcommand{\bA}{ \mbox{\bf A}}

\newcommand{\bP}{ \mbox{\bf P}}

\newcommand{\bX}{ \mbox{\bf X}}

\newcommand{\bZ}{ \mbox{\bf Z}}

\newcommand{\bY}{ \mbox{\bf Y}}

\newcommand{\bs}{ \mbox{\bf s}}

\newcommand{\bu}{ \mbox{\bf u}}

\newcommand{\bV}{ \mbox{\bf V}}

\newcommand{\bI}{ \mbox{\bf I}}

\newcommand{\bM}{ \mbox{\bf M}}

\newcommand{\iid}{\stackrel{iid}{\sim}}

\newcommand{\calR}{{\cal R}}

\newcommand{\Matern}{ \mbox{Mat$\acute{\mbox{e}}$rn}}

\newcommand{\beq}{ \begin{equation}}
\newcommand{\eeq}{ \end{equation}}
\newcommand{\beqn}{ \begin{eqnarray}}
\newcommand{\eeqn}{ \end{eqnarray}}

\usepackage{caption}
\newtheorem{prop}{Proposition}

\begin{document} 

\begin{center}
{\Large Spatial regression modeling via the R2D2 framework}\\\vspace{6pt}
{\large Eric Yanchenko\footnote[1]{North Carolina State University}, Howard D. Bondell\footnote[2]{ University of Melbourne} and Brian J. Reich$^1$}\\
\today
\end{center}

\begin{abstract}\begin{singlespace}\noindent
Spatially dependent data arises in many applications, and Gaussian processes are a popular modelling choice for these scenarios. While Bayesian analyses of these problems have proven to be successful, selecting prior distributions for these complex models remains a difficult task. In this work, we propose a principled approach for setting prior distributions on model variance components by placing a prior distribution on a measure of model fit. In particular, we derive the distribution of the prior coefficient of determination. Placing a beta prior distribution on this measure induces a generalized beta prime prior distribution on the global variance of the linear predictor in the model. This method can also be thought of as shrinking the fit towards the intercept-only (null) model. We derive an efficient Gibbs sampler for the majority of the parameters and use Metropolis-Hasting updates for the others. Finally, the method is applied to a marine protection area data set. We estimate the effect of marine policies on biodiversity and conclude that no-take restrictions lead to a slight increase in biodiversity and that the majority of the variance in the linear predictor comes from the spatial effect.\vspace{12pt}

 \end{singlespace}
\end{abstract}

\newpage

\section{Introduction}\label{s:intro}
Spatially-dependent data arise in many applications including ecology \citep[e.g.,][]{plant2018spatial}, public health 
\citep[e.g.,][]{reich2018precision}, environmental exposure monitoring \citep[e.g.,][]{berrocal2020comparison} and medical imaging \citep[e.g.,][]{masotti2021novel}. 
For example, this work is motivated by an analysis of the effect of conservation efforts on aquatic biodiversity \citep{gill2017capacity}.  Since biodiversity is driven by  ecological processes that evolve over space and time, it is reasonable to expect dependence between nearby observations.  Accounting for this spatial dependence is necessary to provide accurate inference of covariate effects \citep{hodges2010adding} and allow for prediction of biodiversity at unmeasured locations.

Bayesian inference is a popular approach for modeling spatial data. This is typically accomplished through a latent Gaussian process. Estimating the hyper-parameters of this covariance structure is notoriously difficult \citep[e.g.,][]{zhang2004inconsistent} which makes selecting prior distributions paramount. The standard approach places a vague (large variance) inverse gamma prior distribution on the spatial variance and informative gamma prior distribution on spatial correlation parameters \citep{gelfand2010handbook}. As \cite{berger2001objective} discuss, eliciting an informative prior distribution is challenging as the spatial correlation parameters can be difficult to interpret. To provide an automatic approach, the authors derive reference priors for the covariance parameters using an objective Bayesian paradigm which minimizes the prior information from an information theory perspective \citep{berger2006}. The Bayesian hierarchical spatial model is quite intricate so the resulting prior distribution has a complicated form. To provide a prior construction that can more easily incorporate prior knowledge while still being weakly informative, \cite{fuglstad2019constructing} adapt the penalized complexity prior framework of \cite{simpson2017penalising} to construct prior distributions for the hyper-parameters of a Mat\'{e}rn covariance structure. This prior distribution shrinks the spatial component of the model towards a base model, i.e., one with no spatial dependence, and the resulting prior distribution is much faster and easier to compute than that of \cite{berger2001objective}.

In this work, we propose a principled prior framework for Gaussian process spatial models by leveraging 
a Bayesian coefficient of determination, $R^2_n$, \citep{gelmanf2019r} and the R2D2 prior framework \citep{zhang2022bayesian}. This extends \cite{zhang2022bayesian}, \cite{yanchenko2021r2d2} and \cite{aguilar2022intuitive} to spatial models. 
We show that a beta prior distribution on $R^2_n$ is (approximately) equivalent to a conditional generalized beta prime distribution on the linear predictor variance, which includes the marginal spatial variance. This derivation conditions on other spatial parameters which allows the proposed method to accommodate virtually any correlation structure, e.g., non-stationarity. We also derive an efficient MCMC sampler which consists almost entirely of Gibbs sampling steps. The resulting prior specification provides both an interpretable method to select a subjective prior, and a default approach in the absence of prior information. Indeed, the Bayesian coefficient of determination measures the proportion of variance explained by the model so prior domain knowledge can be intuitively incorporated in the model through $R^2_n$ as opposed to choosing prior distributions for individual spatial parameters. For example, it is likely easier for a practitioner to have a prior belief that the model will capture a certain proportion of the variation in the data than to {\it a priori} estimate the spatial range parameter. On the other hand, a uniform prior distribution on $R^2_n$ or one with large prior mass near zero yields an automatic approach. Lastly, we apply the proposed method to a marine protection area data set to study the effects of marine policies and find that certain fishing restrictions lead to a small increase in aquatic biodiversity.

The proposed method relates to the existing methods in several ways. First, each is weakly informative in some sense: \cite{berger2001objective} in terms of information theory, \cite{fuglstad2019constructing} in terms of the Kullback-Leibler divergence between the base and fitted model, and the proposed method (can be) in terms of $R^2_n$. Similar to \cite{fuglstad2019constructing}, the proposed method allows for intuitive incorporation of prior domain knowledge, facilitates fast and easy computation and shrinks the fit towards a base model. Specifically, a prior distribution with large mass near $R^2_n=0$ is akin to shrinking or penalizing towards the intercept-only (baseline) model. The difference is that \cite{fuglstad2019constructing} treats the model without spatial dependence as the baseline model. In contrast with the previous two methods, our approach chooses the prior distribution for the global variance which induces a prior distribution on the other spatial parameters whereas \cite{berger2001objective} and \cite{fuglstad2019constructing} directly set the prior distribution on the spatial hyper-parameters and marginal variance. 

The layout for the remainder of the paper is as follows. In Section \ref{s:methodGP} we present the modeling framework. Section \ref{sec:r2d2} derives the prior distribution and discusses its properties. Computation is the topic of Section \ref{s:comp} and we apply the method to a marine protection area data set in Section \ref{s:data}. We close with a discussion in Section \ref{s:conc}.

\section{Spatial Gaussian process model}\label{s:methodGP}
We introduce the method for a Gaussian spatial regression model with fixed effects. For observation $i\in\{1,\dots,n\}$, let $Y_i$ be the response, $\bs_i\in\calR^2$ be the spatial location and $\bX_i=(X_{i1},\dots,X_{ip})$ be a corresponding vector of explanatory variables. We also define the $n\times p$ matrix $\bX = (\bX_1^T,\dots,\bX_n^T)^T$ and $n\times 2$ matrix $\bs = (\bs_1^T,\dots,\bs_n^T)^T$. The standard spatial regression model is
\begin{equation}\label{e:Ymodel}
  Y_i=\beta_0+\bX_i\bbeta + \theta_i + \varepsilon_i
\end{equation}
where $\beta_0$ is the intercept, $\bbeta=(\beta_1,\dots,\beta_p)^T$ is a vector of fixed effects, $\btheta=(\theta_1,\dots,\theta_n)^T$ are the spatial random effects and $\varepsilon_i\iid\mathsf{Normal}(0,\sigma^2)$ is non-spatial error. We define $\eta_i=\beta_0 + \bX_i\bbeta+\theta_i$ as the linear predictor for observation $i$. 

\subsection{Modeling framework}
The covariance of the fixed and random effects are parameterized in terms of an overall variance parameter $W>0$, spatial correlation parameters $\bpsi$ and proportions $\bphi = (\phi_1,\dots,\phi_{p+1})$ with $\phi_j>0$ and $\sum_{j=1}^{p+1}\phi_j=1$.  The elements of $\bphi$ apportion variance between the fixed and spatial random effects.  Specifically, \begin{equation}\label{e:RE_model}
\bbeta|\sigma^2,W,\phi\sim \mathsf{Normal}({\bf 0}_p,\sigma^2 W\Phi) \mbox{\ \ and \ \ }\btheta|\sigma^2,W,\bphi,\bpsi\sim \mathsf{Normal}({\bf 0}_n,\sigma^2\phi_{p+1}W\Sigma_{\bpsi})
\end{equation}
where ${\bf 0}_m$ is a vector of zeros of length $m$, $\Phi$ is a $p\times p$ diagonal matrix with diagonal elements $\{\phi_1,\dots,\phi_p\}$, $\Sigma_{\bpsi}$ is the $n\times n$ spatial correlation matrix (for notational convenience we will often denote the spatial correlation matrix simply as $\Sigma$), and $\bbeta\perp\btheta$. We stress that the prior construction is {\it conditional} on the spatial correlation matrix $\Sigma$; therefore the framework holds for virtually any correlation function, e.g., Mat\'{e}rn, anisotropic, etc.

The ensuing derivations treat the explanatory variables $\bX$ as fixed. It is common practice to standardize $\bX$ such that its columns have mean zero and variance one, although the results still hold even if this is not the case. We recommend standardizing $\bX$ so that $\sigma^2W$ is the average marginal global variance of the linear predictor $\eta_i$ (see Supplemental Materials). Indeed, $W$ is then the average signal-to-noise ratio, i.e., 
$$
    \frac1n\sum_{i=1}^n\frac{\mathsf{Var}(\eta_i)}{\mathsf{Var}(\varepsilon_i)}
    =\frac{\sigma^2W}{\sigma^2}
    =W.
$$

\subsection{Dirichlet decomposition}

The parameters $\bphi$ determine the relative variance of the model apportioned to each fixed effect and the spatial effect. The proportion of variance allocated to the spatial effect is $\phi_{p+1}$ and $1-\phi_{p+1}=\sum_{j=1}^p\phi_j$ is the proportion of variance allocated to the fixed effects. We could consider these parameters as fixed or give them a prior distribution, e.g., $\bphi\sim \mathsf{Dirichlet}(\xi_1,\dots,\xi_{p+1})$. We often take $\xi_1=\cdots=\xi_{p+1}\equiv\xi_0$ as default choice of these hyper-parameters. The concentration parameter $\xi_0>0$ controls the prior variation with large $\xi_0$ encouraging all the variance components to be roughly equal to $1/(p + 1)$ and small $\xi_0$ encouraging prior uncertainty in the variance components. Another default choice is $\xi_1=\cdots=\xi_p=\frac{1}{2p}$ and $\xi_{p+1}=\frac12$ if it is expected that the fixed and spatial effects contribute equally to the global variance. Lastly, if we believe that the variance is the same for each fixed effect, then we can define a new parameter, $\tilde{\boldsymbol{\phi}}$, such that $\tilde{\boldsymbol\phi}=(\tilde\phi_1,\tilde\phi_2)\sim\mathsf{Dirichlet}(\xi_0,\xi_0)$. Then we set $\phi_j=\tilde\phi_1/p$ for $j\in\{1,\dots,p\}$ and $\phi_{p+1}=\tilde\phi_2$ to enforce equal variance for all fixed effects. We suggest this final parameterization as a default choice due to its simplicity.

\section{Spatial R2D2 prior}\label{sec:r2d2}
The goal of this work is to chose prior distributions that induce a desirable distribution on the Bayesian coefficient of determination, $R^2_n$, from \cite{gelmanf2019r} defined below in Section \ref{sec:R2}. To achieve this, we specify a prior distribution for $W$ given $\bphi$ and $\bpsi$ that induces a Beta$(a,b)$ on $R^2_n$.  Since the prior for $R^2_n$ is $\mathsf{Beta}(a,b)$ for all $\bphi$ and $\bpsi$, the marginal distribution of $R^2_n$ over $\bphi$ and $\bpsi$ is also $\mathsf{Beta}(a,b)$.  In this sense, the prior for $R^2_n$ described below is a function of the joint prior distribution of $W$, $\bphi$ and $\bpsi$.     

\subsection{Bayesian coefficient of determination}\label{sec:R2}

 If we define $\eta_i$ as the signal and $\varepsilon_i$ as the error, then 
$R^2_n$ \citep{gelmanf2019r} is
\begin{equation}\label{e:R2}
 R_n^2 = \frac{v_n}{v_n+\sigma^2}
\end{equation}
where $v_n=\sum_i(\eta_i-\bar\eta)^2/(n-1)$ is the sample variance of the signal and $\bar \eta=\sum_{i=1}^n\eta_i/n$ is the sample mean.\footnote[1]{We do not derive the prior distribution in terms of the {\it marginal} Bayesian $R^2$ as in \cite{zhang2022bayesian} and \cite{yanchenko2021r2d2} because this marginalization removes the effect of the spatial parameters $\bpsi$.} 
The interpretation of  $v_n$ is the variation of the expectation of future data, conditioned on the fixed and spatial effects as well as the explanatory variables and spatial locations. Since $v_n$ is the variance of the modeled predictive means, and the predicted means depend on model parameters, then $v_n$ is conditional on the distribution of the fixed effects. 
Therefore, $R^2_n$ measures model complexity because a more complex model can explain more variation in future data than a simpler model. For example, with $W=0$ implying $\bbeta={\bf0}_p$ and $\btheta={\bf0}_n$, then $\eta_i=\beta_0$ is the intercept-only model and $v_n=R^2_n=0$. Thus, a prior for $W$ with mass near zero corresponds to a prior for $R^2_n$ with mass near zero and shrinks the prior to the simple intercept-only model.

\subsection{Prior derivation}

While \cite{gelmanf2019r} proposed $R_n^2$ as an {\it a posteriori} measure of model fit, we use $R_n^2$ to determine the prior distribution of covariance parameters.  We view $\bX$ and $\bs$ as fixed and thus $\eta_i$ is a random function of $\bbeta$ and $\btheta$, whose distribution depends on $W$, $\bpsi$ and $\bphi$ in (\ref{e:RE_model}).  In this section, we specify a prior for $W$ given $\bpsi$ and $\bphi$ so that averaging over the joint prior distributions $\bbeta$, $\btheta$ and $W$ gives $R^2_n\sim \mathsf{Beta}(a,b)$. 

In order for $R^2_n\sim\mathsf{Beta}(a,b)$, the variance $v_n$ must have {\it generalized beta prime (GBP) distribution} \citep{johnson1995, yanchenko2021r2d2}, i.e., $v_n\sim \mathsf{GBP}(a,b,1,\sigma^2)$.  The GBP distribution can be obtained via a transformation of a beta random variable: if $U\sim\mathsf{Beta}(a,b)$, then $X=d\{U/(1-U)\}^{1/c}\sim\mathsf{GBP}(a,b,c,d)$ and has density function
\begin{equation}\notag
    \pi(x;a,b,c,d)
    =\frac{c\left(\frac{x}{d}\right)^{a c-1}\left(1+\left(\frac xd\right)^c\right)^{-a-b}}{dB(a,b)},\ x\geq0
\end{equation} 
for $a,b,c,d>0$ where $B(\cdot,\cdot)$ is the beta function.  The GBP reduces to the {\it beta prime distribution} if $c=d=1$. 

To specify the prior we write 
\begin{equation}\label{eq:var}
    v_n
    =(\bX\bbeta+\btheta)^T\bP(\bX\bbeta+\btheta),
\end{equation}
where $\bP=(\bI_n-\frac1n{\bf 1}_n{\bf 1}_n^T)/(n-1)$, $\bI_n$ is the $n\times n$ identity matrix and ${\bf 1}_n$ is the vector of ones of length $n$. Since the fixed and spatial effects are assumed independent and we condition on $\bX$, $\bX\bbeta+\btheta$ follows a normal distribution with mean zero and covariance $\sigma^2 W (\bX\Phi\bX^T + \phi_{p+1}\Sigma)$.  Then the distribution of $v_n$ can be equivalently written in terms of $\bZ=(\sigma^2W)^{-1/2}(\bX\bbeta+\btheta)$ where $\bZ\sim\mathsf{Normal}({\bf 0}_n, \bX\Phi\bX^T+\phi_{p+1}\Sigma)$ and
\begin{equation}
    v_n
    =\sigma^2WS
\end{equation}
where $S=\bZ^T\bP\bZ$. Thus, $v_n=\sigma^2WS\sim \mathsf{GBP}(a,b,1,\sigma^2)$ is equivalent to $WS\sim\mathsf{BP}(a,b)$. 

To specify a prior distribution for $W$ so that the product $WS$ has a beta prime distribution requires an understanding of the distribution of $S$. The classical quadratic form results that yield a chi-squared distribution do not apply here because the covariance of $\bZ$ is not idempotent. Thus, the true distribution is a weighted sum of $\chi^2_1$ random variables. Instead of working with the exact distribution, we approximate it by a gamma distribution with the same mean and variance \citep[e.g.,][]{box1954some}. By properties of quadratic forms,
\begin{align*}\notag
    \mathsf{E}(S)
    &=\mu_S=\mbox{tr}(\bP\bX\Phi\bX^T)+\phi_{p+1}\mbox{tr}(\bP\Sigma) \\ \mathsf{Var}(S)
    &=\sigma^2_S=2\mbox{tr}\{\bP(\bX\Phi\bX^T + \phi_{p+1}\Sigma)\bP(\bX\Phi\bX^T + \phi_{p+1}\Sigma)\}.
\end{align*}
So, $S\stackrel{approx.}{\sim}\mathsf{Gamma}(\alpha_{\psi,\phi},\beta_{\psi,\phi})$ where $\alpha_{\psi,\phi}=\mu_S^2/\sigma^2_S$ and $\beta_{\psi,\phi}=\sigma^2_S/\mu_S$ and $\beta_{\psi,\phi}$ is the scale parameter.  

Since $WS\sim\mathsf{BP}(a,b)$ and $S\stackrel{approx.}{\sim}\mathsf{Gamma}(\alpha_{\psi,\phi},\beta_{\psi,\phi})$, then the conditional distribution of $W$ given $\bphi$ and $\bpsi$ is specified by the hierarchical model
\begin{equation}\label{eq:w1}
    W = U_1V
\end{equation}
where $U_1\sim \mathsf{BP}(a,b)$ and $V|\bpsi,\boldsymbol\phi\sim \mathsf{IG}(\alpha_{\psi,\phi},\beta_{\psi,\phi})$ and $\mathsf{IG}(b_1,b_2)$ is the {\it inverse gamma distribution}. This expression shows that the distribution of $W$ is conditional on the spatial and variance allocation parameters, $\bpsi$ and $\boldsymbol\phi$, respectively. This allows trivial extensions of the prior construction to different spatial variance models. The distribution of $R^2_n$, however, is unconditional on any model parameters because of this construction.

The distribution in (\ref{eq:w1}) is not standard so we would like to write it in a more manageable form. We begin with the following proposition.

\begin{prop}\label{prop:1}
If $X|\gamma\sim\mathsf{Gamma}(a,\gamma^{-1})$ and $\gamma\sim\mathsf{Gamma}(b,1)$, then $X\sim\mathsf{BP}(a,b)$.
\end{prop}

\noindent
By Proposition \ref{prop:1}, (\ref{eq:w1}) is equivalent to 
\begin{equation}\label{eq:w2}
    \gamma\sim\mathsf{Gamma}(b,1),\ \ \ \ \ W|\bX,\bs,\boldsymbol\phi,\bpsi,\gamma\sim U_2V
\end{equation}
where $U_2\sim \mathsf{Gamma}(a,\gamma^{-1})$.
We utilize this form of the distribution for computational purposes, but we can further simplify it for better conceptual understanding.

\begin{prop}\label{prop:2}
If $X_1\sim \mathsf{Gamma}(a_1,b_1)$ and $X_2\sim\mathsf{Gamma}(a_2,b_2)$, then $X_1/X_2\sim\mathsf{GBP}(a_1,a_2,1,b_1/b_2)$.
\end{prop}

\noindent
Therefore, an equivalent expression of (\ref{eq:w1}) and (\ref{eq:w2}) is
\begin{equation}\label{eq:w3}
    \gamma\sim\mathsf{Gamma}(b,1),\ \ \ \ \ W|\bX,\bs,\boldsymbol\phi,\bpsi,\gamma\sim\mathsf{GBP}(a,\alpha_{\psi,\phi},1,(\beta_{\psi,\phi}\gamma)^{-1}).
\end{equation}

We can, thus, write out the full prior specification:
\begin{align}
\notag
    &\bbeta|\sigma^2,W,\bphi\sim\mathsf{Normal}({\bf 0}_p,\sigma^2W\Phi),\\\notag &\btheta|\sigma^2,W,\bphi,\bpsi\sim\mathsf{Normal}({\bf 0}_n,\sigma^2\phi_{p+1}W\Sigma),\\ 
    &W|\phi,\bpsi,\gamma\sim\mathsf{GBP}(a,\alpha_{\psi,\phi},1,(\beta_{\psi,\phi}\gamma)^{-1})\\\notag
    &\gamma\sim\mathsf{Gamma}(b,1)\\\notag &\boldsymbol\phi\sim\mathsf{Dirichlet}(\xi_0,\dots,\xi_0).\notag
\end{align}
Taking $\beta_0\sim\mathsf{Normal}(\mu_0,\sigma^2_0),\ \sigma^2\sim\mathsf{IG}(a_0,b_0)$ and $\bpsi\sim\pi(\cdot)$ completes the model. Since this prior specification was induced by a prior distribution on $R^2$ and the variance is apportioned to the fixed and spatial effects through the Dirichlet decomposition, we name this the {\it spatial R2D2 prior} as in \cite{zhang2022bayesian} and \cite{yanchenko2021r2d2}.

\subsection{Properties}
 
We now explore different properties of the spatial R2D2 prior. First, the unconditional distribution (not depending on $\gamma$) of $W$ is not analytic, but we can find its unconditional prior mean and variance.

\begin{prop}
Let $\gamma\sim\mathsf{Gamma}(b,1)$ and $W|\gamma\sim\mathsf{GBP}(a,\alpha_{\psi,\phi},1,(\beta_{\psi,\phi}\gamma)^{-1})$. If $\alpha_{\psi,\phi},b>1$, then
$$
    \mathsf{E}(W) =\frac{a}{\beta_{\psi,\phi}(\alpha_{\psi,\phi}-1)(b-1)}
$$
and $\mathsf{E}(W)=\infty$ otherwise. If $\alpha_{\psi,\phi},b>2$, then
$$
    \mathsf{Var}(W)
    =\frac{a(a+\alpha_{\psi,\phi}-1)}{\beta_{\psi,\phi}^2(\alpha_{\psi,\phi}-2)(\alpha_{\psi,\phi}-1)^2(b-1)(b-2)} + \frac{a^2}{\beta_{\psi,\phi}^2(\alpha_{\psi,\phi}-1)^2(b-1)^2(b-2)}
$$
and $\mathsf{Var}(W)=\infty$ otherwise.
\end{prop}

\noindent
While $\alpha_{\psi,\phi}$ and $\beta_{\psi,\phi}$ are complex functions of the explanatory variables, spatial locations and spatial covariance matrix, $a$ and $b$ can be tuned to enforce certain properties on $W$. For example, the expectation of $W$ increases as $a/b$ increases (which also increases the prior mean of $R^2_n$) and $b>2$ is required for both finite and mean and variance. Conversely, if $b$ is large then there is large prior mass near $R^2_n=0$ and the prior mean of $W$ is also small. If $a<1$ and $b\leq1$, then $W$ has infinite expectation (i.e., heavy tail) as well as a mode at 0, a desirable property of shrinkage priors \citep[e.g.][]{carvalho2009handling, bhattacharya2015, zhang2022bayesian}. 

\begin{figure}
    \centering
    \includegraphics{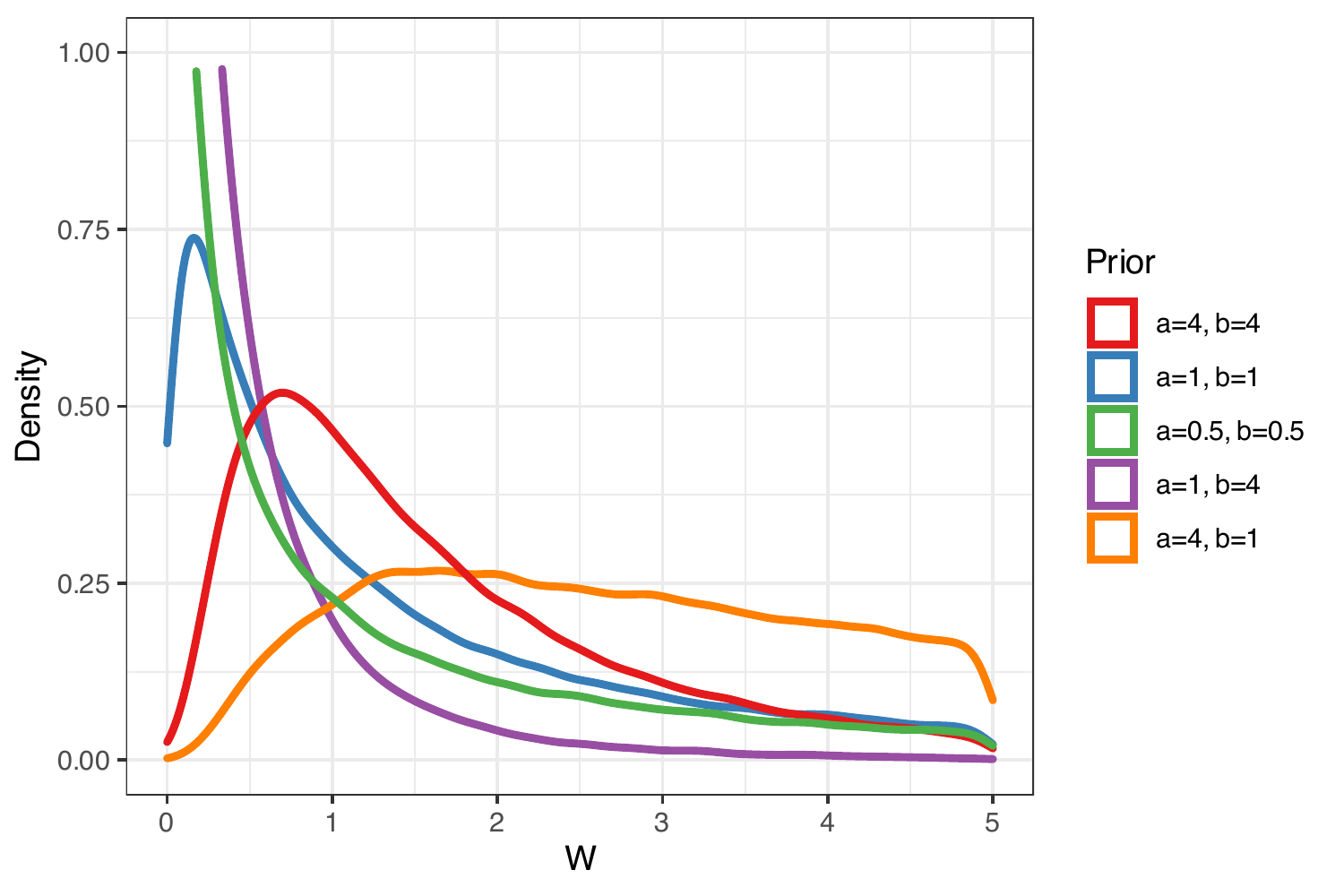}
    \caption{Prior distribution of $W$ for different combinations of $(a,b)$.}
    \label{fig:prior_w}
\end{figure}

We also plot the prior distribution of $W$ under different settings. Let $n=100$, $p=5$, $\bphi=(0.1,0.1,0.1,0.1,0.1,0.5)$ and $X_{ij}\iid \mathsf{Normal}(0,1)$. The locations $\bs_i$ are sampled uniformly from $[0,1]^2$ and we use the Mat\'{e}rn correlation function with $\nu=\rho=0.5$. In Figure \ref{fig:prior_w} we plot the prior distribution of $W$ for $(a,b)\in\{(1,1), (1,4), (4,1), (4,4), (0.5,0.5)\}$ for a particular realization of $\bX$ and $\bs$. For priors with large mass near $R^2_n=0$, e.g.,  $(a,b)=(1,4)$, the prior for $W$ has a mode at $W=0$. On the other hand, the distribution of $W$ for $(a,b)=(4,1)$ has much heavier tails. 

In Table \ref{tab:prior_w} we report the prior mean and variance of $W$ for the same settings as above but let $\rho\in\{0.2,0.4,0.6,0.8,1.0\}$. Again, these results are for a specific realization of covariates and sampling locations. To ensure finite mean and variance of $W$ we take $(a,b)=(4,4)$. As the spatial correlation increases, both the mean and variance of $W$ increase since the model cannot capture the same amount of spatial variation without $W$ increasing. This highlights the interplay between $\rho$ and $W$ for determining the prior distribution.

\begin{table}[]
    \centering
    \begin{tabular}{c|cc|cc}
        $\rho$ & $\alpha_{\psi,\phi}$ & $\beta_{\psi,\phi}$ & $\mathsf{E}(W)$ & $\mathsf{Var}(W)$  \\\hline
        0.2 & 6.63 & 0.14 & 1.70 & 3.68\\
        0.4 & 5.37 & 0.16 & 1.97 & 5.56\\
        0.6 & 4.72 & 0.16 & 2.21 & 7.62\\
        0.8 & 4.32 & 0.17 & 2.40 & 9.72\\
        1.0 & 4.05 & 0.17 & 2.57 & 11.82
    \end{tabular}
    \caption{Values of parameters $\alpha_{\psi,\phi}$ and $\beta_{\psi,\phi}$ and summaries of prior distribution of $W$ for $(a,b)=(4,4)$ and different values of the spatial range parameter $\rho$.}
    \label{tab:prior_w}
\end{table}

Since the prior distribution of $W$ is highly dependent on $\mu_S$ and $\sigma^2_S$, we are interested in their behavior for several specific correlation structures without covariates, i.e., $\phi_{p+1}=1$.  
Recall that the prior for $W$ in (\ref{eq:w1}) is $WS|\bphi,\bpsi\sim\mathsf{BP}(a,b)$ for scaling factor $S$ that is (approximately) distributed as a gamma random variable with mean $\mu_S$ and variance $\sigma_S^2$. This mean and variance depend on the spatial correlation matrix and therefore studying their forms can illustrate how spatial correlation affects $W$'s prior distribution. In the following examples, we consider three specific spatial correlation structures $\Sigma$ and their effect on the prior distribution of $W$.

\paragraph{Example 1 -- Compound symmetry} First consider the compound symmetry model with correlation $\rho$ between all observations. Then 
$$\mu_S = 1-\rho \hspace{20pt}\mbox{and}\hspace{20pt} \sigma_S^2=2\frac{(1-\rho)^2}{n-1}.$$
Therefore, if $\rho<1$ then $S\to\mu_S$ in distribution and, in particular, if the observations are independent with $\rho=0$ then $S$ converges in distribution to one and $W\sim\mathsf{BP}(a,b)$ as in \cite{zhang2022bayesian}. On the other hand, if $\rho=1$ then $S$ is degenerate at zero.  In this extreme, the covariance is singular and restricts $v_n=0$ so that no prior for $W$ can achieve the prior $R_n^2\sim\mathsf{Beta}(a,b)$. For $\rho$ near to but less than one, $S$ has mean and variance near zero implying the prior scale of $W$ must be large to compensate for the restrictions of $v_n$ induced by the correlation.

\paragraph{Example 2 -- Blocked compound symmetry} Next assume the $n$ locations are partitioned into $m$ blocks, each with $n/m$ locations, and the correlation is $\rho$ within block and zero between blocks.  Then 
$$
    \mu_S = 1-\frac{1}{m}\frac{n-m}{n-1}\rho.
$$
We recover the compound symmetry result for $m=1$ and the mean increases as the number of blocks $m$ increases and/or as the correlation decreases. As the number of blocks increases, the number of correlated observations decreases which causes the prior mean of $W$ to decrease since the mean of $W$ and $S$ are inversely proportional. Put another way, a smaller spatial correlation does not require as large of a $W$ to capture the dependency. 

\paragraph{Example 3 -- $\Matern$ correlation} For the general $\Matern$ (or any other) spatial correlation model we have
$$
    \mu_S = 1-\frac{2}{n(n-1)}\sum_{i< j}\rho(d_{ij})
    \to 1-\mathsf{E}\{\rho(D)\}
$$
as $n\to\infty$ where the expectation is with respect to the sampling distribution of the spatial locations\footnote{While in the prior construction we considered the spatial sampling locations to be fixed, for theoretical study it is more convenient to treat them as random.}. If the correlation function is convex (e.g., exponential), then
$$
    \mu_S
    = 1-\frac{2}{n(n-1)}\sum_{i< j}\rho(d_{ij})
    \leq 1-\rho(\bar d)
$$
where $\bar d$ is the average distance between points. Again, we see that $\mu_S$ is inversely related to the strength of spatial correlation.

\section{Posterior computation}\label{s:comp}

The posterior distribution is approximated using a combination of Gibbs and Metropolis-Hastings sampling. We use the formulation of $W$ from Proposition \ref{prop:1}. Specifically, $\gamma\sim\mathsf{Gamma}(b,1)$, $U|\gamma\sim\mathsf{Gamma}(a,\gamma^{-1})$, $V\sim\mathsf{IG}(\alpha_{\psi,\phi},\beta_{\psi,\phi}^{-1})$ and then $W=UV$. Additionally, let $Z\sim\mathsf{giG}(\rho, \chi,\lambda)$ be a generalized inverse Gaussian distribution with density function $p(z)\propto z^{\lambda-1}\exp\{-(\rho z+z/\chi)/2\}$. The MCMC sampler is then as follows.
\begin{enumerate} 
    \item $\beta_0|{\bf Y},{\bf X},\bbeta,\btheta,\sigma^2\sim\mathsf{Normal}(VM,V)$ where $M=\frac1{\sigma^2}{\bf 1}_n^T({\bf Y}-\bX\bbeta-\btheta) + \frac1{\sigma^2_0}\mu_0$ and $V=(n/\sigma^2+1/\sigma^2_0)^{-1}$
    \item $\bbeta|\bY,\bX,\btheta,U,V,\boldsymbol\phi,\sigma^2\sim \mathsf{Normal}(\bV_1\bM_1, \sigma^2\bV)$ where $\bM_1=\bX'(\bY-\beta_0{\bf 1}_n-\boldsymbol\theta)$ and $\bV_1=\{\bX'\bX+ (UV\Phi)^{-1}\}^{-1}$
    \item $\btheta|\bY,\bX,\bbeta,U,V,\boldsymbol\phi,\sigma^2,\bpsi\sim\mathsf{Normal}(\bV_2\bM_2,\sigma^2\bV_2)$ where $\bM_2=(\bY-\beta_0{\bf 1}_n-\bX\bbeta)$ and $\bV_2=\{\bI_n+(\phi_{p+1}UV\Sigma)^{-1}\}^{-1}$
    \item $\sigma^2|\bY,\bbeta,\btheta,U,V,\boldsymbol\phi,\bpsi\sim \mathsf{IG}(a_0+n+p/2, b_0 + \{(\bY-\beta_0{\bf 1}_n-\bX\bbeta-\btheta)'(\bY-\beta_0{\bf 1}_n-\bX\bbeta-\btheta) + \bbeta'(UV\Phi)^{-1}\bbeta + \btheta'(\phi_{p+1}UV\Sigma)^{-1}\btheta\}/2)$
    \item $U|\bbeta,\btheta,V,\boldsymbol\phi,\bpsi,\gamma\sim\mathsf{giG}(2\gamma, \bbeta'(\sigma^2V\Phi)^{-1}\bbeta + \btheta'(\sigma^2\phi_{p+1}V\Sigma)^{-1}\btheta, a-(n+p)/2)$
    \item $V|\bbeta,\btheta,U,\boldsymbol\phi,\bpsi\sim\mathsf{IG}(\alpha_{\psi,\phi}+(n+p)/2, \beta_{\psi,\phi}^{-1}+\tfrac12\{\bbeta'(\sigma^2U\Phi)^{-1}\bbeta + \btheta'(\sigma^2\phi_{p+1}U\Sigma)^{-1}\btheta\})$
    \item $\gamma|U\sim \mathsf{Gamma}(a+b,1+U)$ 
    \item $\bphi|\beta_0,\bbeta,\btheta,U,V,\sigma^2,\bpsi\sim$ Metropolis-Hastings step.
    \item $\bpsi|\beta_0,\bbeta,\btheta,U,V,\sigma^2,\bphi\sim$ Metropolis-Hastings step
\end{enumerate}

Since $\bphi$ and $\bpsi$ are not conjugate, they require a Metropolis-Hastings step to update. Let $\bphi^{(t)}$ be the value of $\bphi$ at step $t$ of the sampler. Then the candidate value $\bphi^*$ is drawn from $\bphi^*\sim\mathsf{Dirichlet}(c_1\bphi^{(t)})$ for some hyper-parameter $c_1$. Using a similar definition and assuming that the spatial correlation model is Mat\'{e}rn with fixed $\nu$, the candidate value $\rho^*$ is drawn from $\log\rho^*\sim\mathsf{Normal}(\log\rho^{(t)},c_2)$ for hyper-parameter $c_2$. Both $c_1$ and $c_2$ are tuned during the burn-in stage to ensure an acceptance rate between 20\% and 50\%.

Although the steps of the algorithm are straightforward to implement, they can be slow for large $n$. The distribution for $W$ depends on $\alpha_{\psi,\phi}$ and $\beta_{\psi,\phi}$ which are functions of $\phi$ and $\psi$. Since $\bphi$ and $\bpsi$ are updated each iteration of the sampler, so too must $\alpha_{\psi,\phi}$ and $\beta_{\psi,\phi}$ be updated. But these terms depend on traces of matrix multiplication so this is a computationally expensive process. This computational burden can be mitigated by the following observation. For some matrix $\bA$ with eigenvalues $\{\lambda_i\}_{i=1}^n$, $\mbox{tr}(\bA)=\sum_{i=1}^n\lambda_i$ and $\mbox{tr}(\bA^2)=\sum_{i=1}^n\lambda_i^2$. Thus, we can replace the computations of the trace of squared matrices with the sum of eigenvalues. Since we only require the eigenvalues (and not the eigenvectors), this can be done in $O(n^2)$ as compared to the $O(n^3)$ needed for matrix multiplication. Additionally, these eigenvalues only need to be computed once per iteration since $\mu_S\propto\mbox{tr}(\bA)$ and $\sigma^2_S\propto\mbox{tr}(\bA^2)$ where $\bA=\bP(\bX\Phi\bX^T+\phi_{p+1}\Sigma)$.

In the Supplemental Materials, we conduct a small simulation study to compare the proposed R2D2 prior framework with a standard vague prior and the Penalized Complexity (PC) prior of \cite{fuglstad2019constructing}. Under these settings, the R2D2 prior yielded the best estimates of the spatial marginal variance parameter, while all methods performed comparably for estimating the fixed effects and spatial range parameter.

\section{Marine protection area data analysis}\label{s:data}
\subsection{Data and model}
\begin{figure}
    \centering
    \includegraphics[scale=0.99]{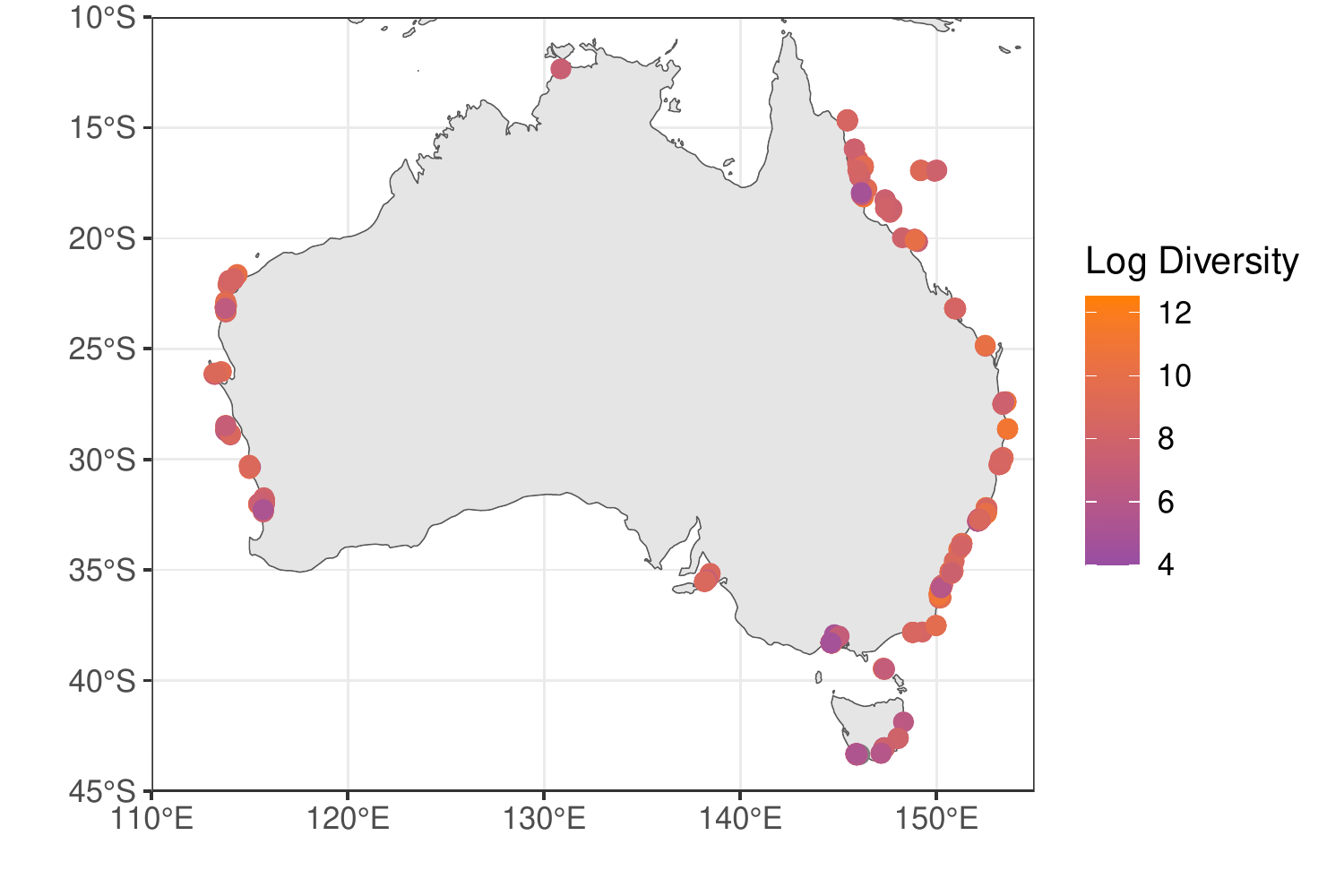}
    \caption{Logarithm of the biodiversity for locations around Australia.}
    \label{fig:aus}
\end{figure}

Marine Protection Areas (MPAs) have been established around the globe to preserve aquatic biodiversity. \cite{gill2017capacity} collected data to understand the effects of these policies on conservation efforts. The response variable, $Y_i$, is the logarithm of the biodiversity at site $\bs_i$ where a larger value of the response means greater biodiversity, a goal of conservationists and scientists. For this analysis, we consider the observations around Australia, as seen in Figure \ref{fig:aus}. Australia is known for its vast and unique collection of species \citep[e.g.,][]{butler2010marine} as well as being home to two {\it biodiversity hotspots} on the east and southwest coasts. Biodiversity hotspots are geographical regions that are rich in species, particularly those that are endemic, rare and/or endangered  \citep{myers1988threatened,reid1998biodiversity}.

The spatial locations of the $n=471$ observations shifted and scaled to fit in the unit square. We also select $p=9$ explanatory variables: multi-use (0) vs. no-take (1) regulation indicator; depth (m); wave exposure (kW/m); distance to shoreline (km); distance to provincial capital market (km); coastal population (within 100 km$^2$); minimum sea surface temperature (2002-2009, $^\circ$C); Chlorophyll-a (2002-2009, mg/m$^3$); and reef area within 15 km. These explanatory variables are centered and scaled to ensure each column has mean zero and variance one. Of primary interest is the indicator variable for whether the sampled location was under a multi-use (MU) or no-take (NT) restriction. MU regions have restrictions on fishing practices but still allow for some fishing whereas NT zones have a total ban on fishing. One question is whether there is a significant difference in biodiversity between the MU and NT zones after accounting for the other covariates.

In addition to these covariates and the spatial random effect, we also include a random intercept for the MPA, which is straightforward to incorporate into our prior framework. The model is then
$$
    Y_i = \beta_0+\bX_i\bbeta+\bZ_i\bu+\theta_i+\varepsilon_i
$$
where $Z_{i\ell}=1$ if site $i$ corresponds with MPA region $\ell$ and 0 otherwise for $\ell\in\{1,\dots,L=37\}$. For the R2D2 prior, the full model specification is
\begin{multline}
    \bbeta|\sigma^2,\phi_1,W\sim\mathsf{Normal}({\bf 0}_p, \tfrac1p\sigma^2\phi_1W\bI_p),\ \ \bu|\sigma^2,\phi_2,W\sim\mathsf{Normal}({\bf 0}_L, \sigma^2\phi_2W\bI_L),\\ \btheta|\sigma^2,\phi_3,W,\rho\sim\mathsf{Normal}({\bf 0}_n, \sigma^2\phi_3W\Sigma),\ \ \gamma\sim\mathsf{Gamma}(b,1),\ \  W|\rho,\phi,\gamma\sim\mathsf{GBP}(a,\alpha_{\psi,\phi},1,(\beta_{\psi,\phi}\gamma)^{-1}).
\end{multline}
where $\Sigma$ is modeled with an exponential correlation structure, i.e.,
$$
    \Sigma_{ij}= e^{-d_{ij}/\rho}
$$
where $d_{ij}=||\bs_i-\bs_j||_2$ is the distance between the sampling locations of observations $i$ and $j$, and $\bpsi=\rho$ is the spatial range parameter. The exponential model is a special case of the Mat\'{e}rn correlation structure with smoothness parameter $\nu=1/2$ \citep{stein1999interpolation}. To complete the prior specification, we let $\beta_0\sim\mbox{Normal}(0,100)$, $\boldsymbol\phi=(\phi_1,\phi_2,\phi_3)\sim\mathsf{Dirichlet}(1,1,1)$, $\sigma^2\sim\mathsf{IG}(0.10,0.10)$ and $\log(\rho)\sim\mathsf{Normal}(-2,1)$.
For computing the hyperparameters $\alpha_{\psi,\phi}$ and $\beta_{\psi,\phi}$, we have $\mu_S=\mbox{tr}(\bA)$ and $\sigma^2_S=2\mbox{tr}(\bA^2)$ where
$$
    \bA
    =\bP(\tfrac{\phi_1}{p}\bX\bX^T + \phi_2\bZ\bZ^T+\phi_3\Sigma).
$$
We consider several version of the spatial R2D2 prior with $(a,b)\in\{(0.5, 0.5),\ (1,1),\ (4,1),\ (1,4)\}$ to understand the effects of the $R^2_n$ prior distribution on the results. We found that the results were mostly insensitive to hyper-parameter choices and report a brief sensitivity analysis in the Supplemental Materials.

For comparison, we also consider a vague prior and the penalized complexity (PC) prior \citep{fuglstad2019constructing}. For the vague prior, we take 
\begin{multline}\notag
    \bbeta\sim\mathsf{Normal}({\bf 0}_p, \sigma^2_\beta\bI_p),\ \  \bu|\sigma^2,\sigma^2_u\sim\mathsf{Normal}({\bf 0}_L, \sigma^2\sigma^2_u\bI_L),\\ \btheta|\sigma^2,\sigma^2_\theta,\rho\sim\mathsf{Normal}({\bf 0}_n, \sigma^2\sigma^2_\theta\Sigma),\ \ \sigma^2_u,\sigma^2_\theta\sim\mathsf{IG}(0.10, 0.10)
\end{multline}
where we fix $\sigma^2_\beta=100$ and the rest of the parameters have the same prior distribution as in the spatial R2D2 prior. From \cite{fuglstad2019constructing}, the PC prior is
\begin{multline}\notag
    \bbeta\sim\mathsf{Normal}({\bf 0}_p, \sigma^2_\beta\bI_p),\  \bu|\sigma^2,\sigma^2_u\sim\mathsf{Normal}({\bf 0}_L, \sigma^2\sigma^2_u\bI_L),\ \btheta|\sigma^2,\sigma^2_\theta,\rho\sim\mathsf{Normal}({\bf 0}_n, \sigma^2\sigma^2_\theta\Sigma),\\ 
    \sigma^2_u\sim\mathsf{IG}(0.10, 0.10),\ \
    \sigma_\theta\sim\mathsf{Exp}(-\log(\alpha)/\sigma_0),\ \
    \rho\sim\mathsf{IG}(1, -\log(\alpha)\rho_0)
\end{multline}
where we set $\alpha=0.05$, $\sigma_0=10$ and $\rho_0=0.01$ as per \cite{fuglstad2019constructing}. Note that these hyper-parameter choices ensure that $\mathsf{P}(\rho<\rho_0)=\alpha$ and $\mathsf{P}(\sigma_\theta > \sigma_0)=\alpha$. Again, all other parameters have the same prior distributions as above.
For all models, we take 10,000 MCMC samples as burn-in, 100,000 to monitor convergence and thin the results by saving every fifth sample. To allow for a fair comparison, we coded all methods by hand in \texttt{R}. The code for the R2D2 prior is available at: \url{https://github.com/eyanchenko/r2d2space}.

\subsection{Results}

We report the posterior median and 95\% credible interval for several parameters of interest in Table \ref{tab:mpa} as well as plot the posterior distribution of $R^2_n$ and $W$ in Figure \ref{fig:mpa}. In the Supplemental Materials, we also report representative sample of trace plots in addition to results on computation time and number of effective samples for each MCMC chain.

\begin{figure}
    \centering
    \includegraphics[scale=0.80]{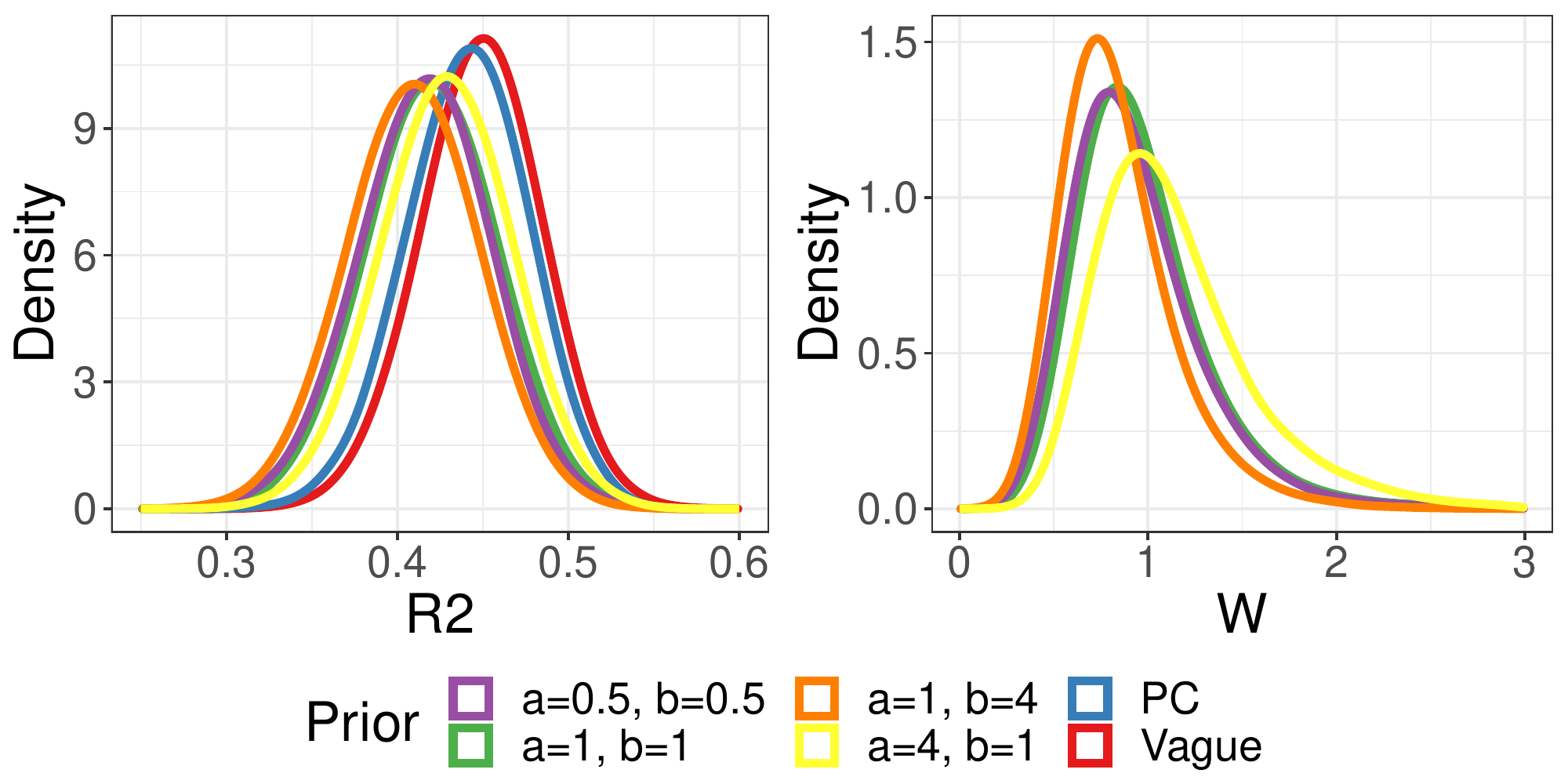}
\caption{Posterior distribution of $R^2_n$ and $W$ for MPA fisheries data using R2D2 prior with $R^2_n\sim\mathsf{Beta}(a,b)$.}
\label{fig:mpa}
\end{figure}

\begin{table}[h]
    \centering
 \begin{tabular}{c|cccc}
        \hline
         $(a,b)$ & $R^2_n$ & $\beta_1$ & $W$ & $\sigma^2_\theta$ \\\hline
         Vague & 0.44 (0.37, 0.51) & 0.11 (-0.03, 0.24) & $-$ & 1.43 (0.36, 7.20) \\
         PC & 0.43 (0.36, 0.50) & 0.10 (-0.03, 0.24) & $-$ & 0.97 (0.25, 4.06)\\
        $(1,1)$ & 0.42 (0.35, 0.50) & 0.07 (-0.05, 0.20) & 0.92 (0.48, 1.88) & 0.56 (0.26, 1.30)\\
        $(\tfrac12,\tfrac12)$& 0.42 (0.35, 0.50) & 0.07 (-0.05, 0.20) & 0.90 (0.47, 1.93) & 0.53 (0.23, 1.26)\\
        $(1,4)$ &  0.41 (0.34, 0.48) & 0.07 (-0.05, 0.19) & 0.80 (0.42, 1.58) & 0.47 (0.18, 1.07) \\
        $(4,1)$ & 0.43 (0.35, 0.50) & 0.07 (-0.05, 0.20) & 1.05 (0.55, 2.10) & 0.60 (0.26, 1.44)\\\hline
        $(a,b)$ & $\rho$ & $\phi_1$ & $\phi_2$ & $\phi_3$ \\\hline
        Vague & 0.23 (0.02, 1.20) & $-$ & $-$ & $-$\\
        PC & 0.18 (0.06, 0.88)  & $-$ & $-$ & $-$\\
        $(1,1)$ & 0.07 (0.01, 0.36) & 0.17 (0.04, 0.44) & 0.17 (0.04, 0.42) & 0.63 (0.36, 0.86)\\
        $(\tfrac12,\tfrac12)$&0.06 (0.01, 0.33) & 0.18 (0.04, 0.50) & 0.18 (0.04, 0.45) & 0.61 (0.30, 0.85)\\
        $(1,4)$ & 0.07 (0.01, 0.34) & 0.17 (0.04, 0.44) & 0.19 (0.04, 0.47) & 0.61 (0.30, 0.85)\\
        $(4,1)$ & 0.07 (0.01, 0.35) & 0.19 (0.04, 0.49) & 0.18 (0.04, 0.44) & 0.60 (0.30, 0.86)
    \end{tabular}
    \caption{Posterior median and 95\% credible intervals for vague, penalized complexity and R2D2 prior with $R^2_n\sim\mathsf{Beta}(a,b)$ for the MPA fisheries data. Note that $\beta_1$ is the effect of NT relative to MU; $\sigma^2_\theta$ is the variance of the spatial term, $\btheta$ ($\phi_3W$ for the R2D2 priors); $\rho$ is the spatial range effect; and $\phi_1,\phi_2$ and $\phi_3$ are the proportion of variance allocated to the fixed, random and spatial effects, respectively.}
    \label{tab:mpa}
\end{table}

First, we notice that each spatial R2D2 prior yields a posterior median of $R^2_n$ around 0.42. This means that approximately 42\% of the variation in the response can be explained by variation in the linear predictor. The posterior median of $R^2_n$ is larger for the $(a,b)=(4,1)$ prior and smaller for the $(a,b)=(1,4)$ prior which reflects their prior means. Interestingly, even though the $(a,b)=(1,1)$ and $(a,b)=(\frac12,\frac12)$ priors have quite different prior shapes, they yield almost identical posterior $R^2_n$ shapes, likely because they have the same prior mean $R^2_n$. As expected, these trends are similar for the posterior distribution of $W$ because $R^2_n$ and $W$ are positively associated. The posterior of $R^2_n$ for the vague and PC priors, however, are slightly larger than that of the spatial R2D2 priors. This is because these priors induce a prior distribution on $R^2_n$ that is effectively a point mass at one, due to the large prior variance of the fixed effects.


Next, we consider the posterior distribution of $\beta_1$, the parameter that quantifies the effect of the NT and MU zones. All models yield a posterior median of $\beta_1$ greater than zero indicating that there is greater biodiversity in the NT zones than in the MU zones. All of the credible intervals contain zero, however, so this finding is not statistically significant. These conclusions are similar to those found in \cite{cui2021advanced}. We can also compute the posterior probability of $\beta_1$ being positive as a measure of significance for this explanatory variable. We find that $\mathsf{P}(\beta_1>0|\bY)$ is 0.94 for the Vague and PC priors compared to 0.88, 0.88, 0.87 and 0.89 for the R2D2 prior with $(a,b)$ equal (1,1), $(\frac12,\frac12)$, (1,4) and (4,1), respectively. This provides moderately strong evidence in favor of a statistically significant effect with the vague and PC priors providing the strongest evidence. Further investigation is needed to understand the practical significance of these zones. 

We also report the posterior median and 95\% credible intervals for the other fixed effects in the Supplemental Materials. For example, sea temperature and measurement depth seem to have the greatest effect on biodiversity whereas population and Chlorophyll-a concentration have very little impact. In many cases, the posterior medians are larger in magnitude for the vague and PC priors because the R2D2 priors are shrinking the fixed effects towards the base model while the prior variance is much larger in the other two models.

The prior distribution of $R^2_n$ has only a small effect on the posterior distribution of $\boldsymbol\phi$ since these distributions are quite similar across different combinations of $(a,b)$. This is sensible because the $R^2_n$ metric is most directly related to the linear predictor, which these parameters have minimal effect on. In particular, the spatial R2D2 prior with $(a,b)=(1,4)$ yields a posterior median of $\boldsymbol\phi=(0.17, 0.19, 0.61)$. Thus, the spatial random effect accounts for approximately 61\% of the variation in the linear predictor, whereas the fixed effects and MPA region account for about 20\% each. The results are similar for the other R2D2 priors.

The difference among the prior frameworks is most pronounced for the spatial range and variance parameters. For the spatial marginal variance $\sigma^2_\theta$ ($\phi_3W$ for the R2D2 priors), the Vague prior has the largest posterior median, followed by the PC prior, while the R2D2 priors have the smallest. The Vague prior does not penalize this parameter which leads to the largest values. Both the PC and R2D2 priors, however, explicitly shrink this parameter towards zero (via $W$ for R2D2), leading to the smaller posterior estimates. From this example, the R2D2 prior corresponds to greater shrinkage. While the R2D2 priors all yield similar results for $\rho$, the PC and vague priors have a posterior median that is approximately three times larger than that of the R2D2 prior. We expect the PC prior to have a larger estimate for $\rho$ because this framework explicitly forces the estimate towards a base model of $\rho\to\infty$. The large estimate for the Vague prior could be owed to the strong correlation (0.68) between the posterior distributions of $\sigma^2_\theta$ and $\rho$. Thus, the large values of $\sigma^2_\theta$ may also be increasing the estimates of $\rho$. The corresponding correlation for the R2D2 $(1,1)$ prior, on the other hand, is only 0.29.

\begin{figure}
    \centering
    \includegraphics[scale=0.75]{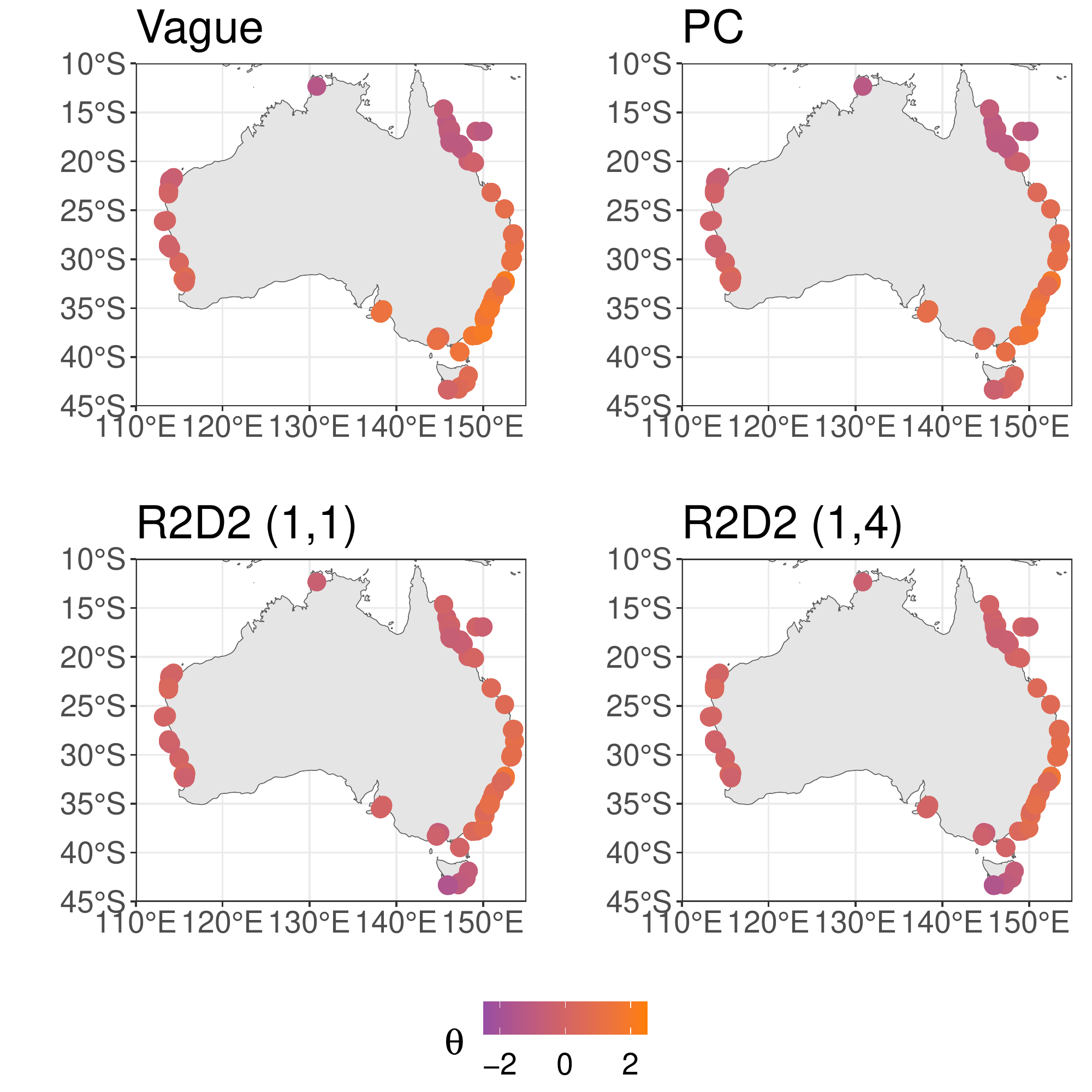}
    \caption{Posterior median of latent spatial parameter $\btheta$ for various prior distributions.}
    \label{fig:theta}
\end{figure}

Finally, in Figure \ref{fig:theta}, we plot the posterior median of $\theta_i$ at each location $\bs_i$. All spatial R2D2 priors gave similar results so we only plot the results for $(a,b)=(1,1)$ and $(1,4)$. From the figure, we notice that there is much greater spatial variability resulting from the vague prior as compared to the R2D2 prior, with the PC prior in between.

\section{Conclusion}\label{s:conc}
In this work, we proposed a novel, principled framework for constructing prior distributions for Gaussian process spatial regression models. The spatial R2D2 prior facilitates an intuitive and interpretable way to incorporate prior information into the statistical model via the Bayesian coefficient of determination. In the absence of prior domain knowledge, we suggest the $(a,b)=(1,1)$ as a natural choice or $(a,b)=(1,b)$ for large $b$ if it is believed that there is sparsity in the fixed effects. Indeed, prior distributions with large mass near $R^2_n=0$ are sensible for variable selection contexts \citep{zhang2022bayesian, yanchenko2021r2d2} which highlights a key connection between the spatial R2D2 prior construction and \cite{fuglstad2019constructing}. \cite{fuglstad2019constructing} shrink towards a ``null model" with $\rho\to\infty$ and $\phi_{p+1}W\to 0$, using the notation of our paper. On the other hand, we consider the intercept-only model with $R^2_n=0$ $(W=0)$ as the baseline model. Thus, a large prior mass near $R^2_n=0$ shrinks towards the null model which is equivalent to a large prior mass of $W$ near zero. This also means that our null model shrinks the fixed effects to zero in addition to the spatial effects. 

We again stress that our prior construction is dependent on the explanatory variables and spatial design. This means that if a new response is observed, then the prior distribution for $W$ would change, similar to \cite{berger2001objective}. The prior distribution in \cite{fuglstad2019constructing}, conversely, is independent of covariates and sampling. While it is important for the practitioners to keep this in mind, from a Bayesian philosophical sense, this is not problematic as results are always thought of as {\it conditional} on the observed data. Additionally, while the proposed prior construction does not account for confounding between the covariates and spatial effect, our method could be used in model-based spatial causal inference analyses, such as those reviewed in \cite{reich2021review}.

It is also straightforward to adapt our framework to other spatial settings. For example, consider a model where the fixed effects are allowed to vary at each spatial location. In particular, let $p=1$ and $\bbeta=(\beta_1,\dots,\beta_n)$ where $\beta_i$ is the fixed effect for spatial location $\bs_i$. To account for the the spatial correlation between the fixed effects, we let $\bbeta|\sigma^2,W,\boldsymbol\phi,\bpsi_{\beta}\sim\mathsf{Normal}({\bf 0}_n, \sigma^2\phi_1W \Sigma_\beta)$ where $\Sigma_\beta$ is a spatial correlation matrix. Then $\alpha_{\psi,\phi}$ and $\beta_{\psi,\phi}$ are computed using $\mu_S=\mbox{tr}(\bA)$ and $\sigma^2_S=2\mbox{tr}(\bA^2)$ where
$$
    \bA
    =\phi_1\bP\bX\Sigma_\beta\bX^T + \phi_2\bP\Sigma.
$$

Finally, there are several interesting avenues of future study for the R2D2 prior framework including areal spatial data, non-Gaussian responses, times series and longitudinal data. In the Supplemental Materials, we outline how this procedure may work for non-Gaussian responses. Additionally, throughout this work we have assumed that $R^2_n$ has a beta distribution {\it a priori}. Another interesting extension would be to consider other prior distributions on $R^2_n$ such as the Kumaraswamy distribution \citep{jones2009kumaraswamy}.

\section*{Acknowledgements}
The authors thank the National Science Foundation (DMS2152887) and the National Institutes of Health (R01ES031651-01) for supporting this work. We would also like the thank the reviewers and editors whose comments greatly improved the quality of the manuscript.

\bibliographystyle{rss}
\bibliography{refs, refs2}

\clearpage

\begin{center}
{\Large Supplementary Materials}
\end{center}

\section{Technical proofs}

\subsection*{Marginal global variance of $\eta_i$}
Assume that $\bX$ is standardized to have columns with mean zero and variance one, i.e., $\sum_{i=1}^n x_{ij}=0$ and $\sum_{i=1}^n x_{ij}^2=n-1\approx n$ for all $j$. Then
\begin{align*}
    \frac1{n}\sum_{i=1}^n \mathsf{Var}(\eta_i)
    &= \frac1{n}\sum_{i=1}^n \mathsf{Var}\left(\beta_0+\sum_{j=1}^p x_{ij}\beta_j +\theta_i\right)\\
    &=\frac1{n}\sum_{i=1}^n \sum_{j=1}^p (x_{ij}^2 \phi_j\sigma^2 W + \phi_{p+1} \sigma^2 W) \\
    &=\left(\sum_{j=1}^p \phi_j\sigma^2 W\frac1n\sum_{i=1}^n x_{ij}^2\right) + \phi_{p+1}\sigma^2 W\\
    &=\sigma^2W\sum_{j=1}^{p+1}\phi_j \\
    &=\sigma^2 W. \ \square
\end{align*}

\subsection*{Proof of Proposition 1}
{\it If $X|\gamma\sim\mathsf{Gamma}(a,\gamma^{-1})$ and $\gamma\sim\mathsf{Gamma}(b,1)$, then $X\sim\mathsf{BP}(a,b)$.}\\
{\it Proof.}
\begin{align*}
    f(x)
    &=\int_{\gamma=0}^\infty f(x|\gamma)f(\gamma)\ d\gamma
    =\int_{\gamma=0}^\infty \frac{\gamma^a}{\Gamma(a)}x^{a-1}e^{-x\gamma}\cdot \frac{1}{\Gamma(b)}\gamma^{b-1}e^{-\gamma} d\gamma\\
    &=\frac1{\Gamma(a)\Gamma(b)}x^{a-1}\int_{\gamma=0}^\infty \gamma^{a+b-1}e^{-(x+1)\gamma}d\gamma
    =\frac1{\Gamma(a)\Gamma(b)}x^{a-1} \frac{\Gamma(a+b)}{(x+1)^{a+b}}\\
    &\sim\mathsf{BP}(a,b)\ \square
\end{align*}

\subsection*{Proof of Proposition 2} {\it If $X_1\sim\mathsf{Gamma}(a_1,b_1)$ and $X_2\sim\mathsf{Gamma}(a_2,b_2)$, then $X_1/X_2\sim\mathsf{GBP}(a_1,a_2,1,b_1/b_2)$.}\\
{\it Proof.} Let $Y_1=X_1/X_2$ and $Y_2=X_2$. Then $X_1=Y_1Y_2$ and $X_2=Y_2$ and $|J|=Y_2$.
Thus,
$$
    f_{(Y_1,Y_2)}(y_1,y_2)
    =f_{X_1}(y_1y_2)f_{X_2}(y_2)\times y_2
    = \frac{1}{\Gamma(a_1)b_1^{a_1} \Gamma(a_2)b_2^{a_2}} (y_1y_2)^{a_1-1} e^{-y_1y_2/b_1} y_2^{a_2} e^{-y_2/b_2},\ y_1,y_2>0.
$$
Therefore,
$$
    f_{Y_1}(y_1)
    \propto y_1^{a_1-1}\int_{y_2=0}^\infty y_2^{a_1+a_2-1} e^{-y_2(y_1/b_1+1/b_2)} dy_2\\
    \propto y_1^{a_1-1} \frac{1}{\{1+y_1/(b_1/b_2)\}^{a_1+a_2}}
    \sim \mathsf{GBP}(a_1,a_2,1,b_1/b_2).
$$

\subsection*{Proof of Proposition 3}
{\it Let $\gamma\sim\mathsf{Gamma}(b,10$ and $W|\gamma\sim\mathsf{GBP}(a,\alpha,1,(\beta\gamma)^{-1}$. If $\alpha,b>1$, then $\mathsf{E}(W)=a/\{\beta(\alpha-1)(b-1)\}$. If $\alpha,b>2$, then}
$$
    \mathsf{Var}(W)
    =\frac{a(a+\alpha-1)}{\beta^2(\alpha-2)(\alpha-1)^2(b-1)(b-2)} + \frac{a^2}{\beta^2(\alpha-1)^2(b-1)^2(b-2)}.
$$
{\it Proof.}
\begin{align*}
    \mathsf{E}(W)
    &=\mathsf{E}_\gamma(\mathsf{E}(W|\gamma))
    =\mathsf{E}_\gamma\left(\frac{a}{(\alpha-1)\beta\gamma}\right)
    =\frac{a}{\beta(\alpha-1)(b-1)}
\end{align*}
where the first equality holds only if $\alpha>1$ and the second requires $b>1$. The variance is found in a similar way noting that
$$
    \mathsf{Var}(W)
    =\mathsf{E}_\gamma(\mathsf{Var}(W|\gamma)) + \mathsf{Var}_\gamma(\mathsf{E}(W|\gamma))
$$
and that the (finite) variance of a GBP holds when $\alpha>2$.

\section{Extension to non-Gaussian response}
We include a brief note on how the R2D2 spatial prior framework may work for non-Gaussian response. As an example of non-Gaussian data, consider a spatial counts regression model where $Y_i\sim\mathsf{Poisson}\{\exp(\bX_i\bbeta+\theta_i)\}$. Let $\mu_i=\exp(\bX_i\bbeta+\theta_i)$ and $\boldsymbol{\mu}=(\mu_1,\dots,\mu_n)^T$. Then we can show that
$$
        R^2_n
        =\frac{\boldsymbol{\mu}^T\bP\boldsymbol{\mu}}{\boldsymbol{\mu}^T\bP\boldsymbol{\mu} + \frac1n{\bf 1}^T\boldsymbol{\mu}}.
$$
where $\bP=(\bI_n-\frac1n{\bf 1}_n{\bf 1}_n^T)/(n-1)$. Now, $\boldsymbol{\mu}$ has a multivariate Log Normal distribution so we cannot use the same techniques from this manuscript. One approach is using simulation to approximately match the prior distribution of $R^2_n$ to the desired $\mathsf{Beta}(a,b)$. For example, we may assume some prior distribution on $(\sigma^2,\boldsymbol{\phi}, \bpsi)\sim\pi(\cdot)$ and that $W\sim\mathsf{GBP}(\alpha_1,\alpha_2,1,\alpha_3)$. We draw from these prior distributions, and find the empirical distribution of $R^2_n$. Then we find the values of $(\alpha_1,\alpha_2,\alpha_3)$ which minimize the KL divergence between the simulated empirical distribution of $R^2_n$ and a desired $\mathsf{Beta}(a,b)$ distribution. This should then yield (approximately) the correct distribution. We may also be able to draw on ideas from Stochastic Variational Inference \citep{hoffman2013stochastic}.

\section{Simulation study}
We conduct a brief simulation study to compare the proposed R2D2 prior framework with that of a vague/uninformative prior and the penalized complexity (PC) prior \citep{fuglstad2019constructing}.

\subsection{Data generation model}
We generate the response, $Y_i$, from a normal likelihood:
$$
    Y_i = \beta_0+\bX_i\bbeta+\theta_i+\varepsilon_i
$$
where $\bbeta\in\mathcal R^p$ are the fixed effects, $\theta_i$ is the spatial effect of response $i$ and $\varepsilon_i\stackrel{iid}{\sim}\mathsf{Normal}(0,\sigma^2)$ for $i=1,\dots,n$. We generate the fixed effects with AR(1) auto-correlation i.e., 
$$
    {\bf X}_i\sim\mathsf{Normal}({\bf 0}_p, \Sigma_X)
$$
where $\Sigma_X$ has an AR(1) auto-correlation, i.e., $\mathsf{Cor}(X_{ij}, X_{ik})=r^{|j-k|}$ for $j\neq k$ and $r\in(0,1)$. Additionally, we generate the true value of the fixed effects from $\beta_j\sim\mathsf{Normal}(0,\sigma^2\sigma^2_\beta)$ for $j=1,\dots,p$. Next, $\boldsymbol{\theta}\sim\mathsf{Normal}({\bf 0}_n,\sigma^2\sigma^2_\theta \Sigma)$ where $\Sigma$ has an exponential correlation structure, i.e.,
$$
    \Sigma_{ij}=e^{-d_{ij}/\rho}
$$
where $d_{ij}=||\bs_i-\bs_j||$ is the distance between the sampling locations of observations $i$ and $j$, and $\rho$ is the spatial range parameter. We generate sampling locations, $\bs_i$, randomly within the unit square.

\subsection{Prior specifications}
We compare the proposed R2D2 prior with a vague prior and PC prior. For the R2D2 prior, the full model specification is
\begin{multline}
    \beta_j|\sigma^2,\phi_j,W\sim\mathsf{Normal}(0, \sigma^2\phi_j W), j=1,\dots,p,\ \ \ \btheta|\sigma^2,\phi_{p+1},W,\rho\sim\mathsf{Normal}({\bf 0}_n, \sigma^2\phi_{p+1}W\Sigma),\ \\ \gamma\sim\mathsf{Gamma}(b,1),\ \  W|\rho,\phi,\gamma\sim\mathsf{GBP}(a,\alpha_{\psi,\phi},1,(\beta_{\psi,\phi}\gamma)^{-1}).
\end{multline}
To complete the prior specification, we let $\beta_0\sim\mbox{Normal}(0,100)$, $\boldsymbol\phi=(\phi_1,\dots,\phi_{p+1})\sim\mathsf{Dirichlet}(1,\dots,1)$, $\sigma^2\sim\mathsf{IG}(0.10,0.10)$ and $\log(\rho)\sim\mathsf{Normal}(-2,1)$. We consider $(a,b)=(1,1), (1,4)$ and $(4,1)$.

For the vague prior, we take 
\begin{multline}\notag
    \bbeta\sim\mathsf{Normal}({\bf 0}_p, \sigma^2_\beta\bI_p),\ \  \btheta|\sigma^2,\sigma^2_\theta,\rho\sim\mathsf{Normal}({\bf 0}_n, \sigma^2\sigma^2_\theta\Sigma),\ \ \sigma^2_\theta\sim\mathsf{IG}(0.10, 0.10)
\end{multline}
where we fix $\sigma^2_\beta=100$ and the rest of the parameters have the same prior distribution as in the spatial R2D2 prior. From \cite{fuglstad2019constructing}, the PC prior is
\begin{multline}\notag
    \bbeta\sim\mathsf{Normal}({\bf 0}_p, \sigma^2_\beta\bI_p),\  \btheta|\sigma^2,\sigma^2_\theta,\rho\sim\mathsf{Normal}({\bf 0}_n, \sigma^2\sigma^2_\theta\Sigma),\\
    \sigma_\theta\sim\mathsf{Exp}(-\log(\alpha)/\sigma_0),\ \
    \rho\sim\mathsf{IG}(1, -\log(\alpha)\rho_0)
\end{multline}
where we set $\alpha=0.05$, $\sigma_0=10\sigma^*_\theta$ and $\rho_0=\rho^*/10$ as per \cite{fuglstad2019constructing}, where $\sigma^*_\theta$ and $\rho^*_\theta$ are the true values of the spatial variance and range, respectively. Again, all other parameters have the same prior distributions as above.

\subsection{Settings and metrics}
For data generation, we set $p=10$ and $r=0.8$. We also choose $\sigma^2=1$, $\sigma^2_\beta=\sigma^2_\theta=0.5^2$ and vary $n=100,200$ and $\rho=0.1, 0.2, 0.3$ for a total of six simulation settings. This leads to a true $R^2_n\approx 0.67$. We generate 11,000 MCMC samples and discard the first 1,000 as burn-in. We compute the root mean squared error (MSE) loss between the true value and posterior median for $\bbeta$, $\sigma^2_\theta$, $\rho$, $R^2_n$ as well as the frequentist coverage of the 95\% credible intervals. Note that the true $R^2_n$ is calculated as $\mathsf{Var}(\eta_1,\dots,\eta_n)/\{\mathsf{Var}(\eta_1,\dots,\eta_n) + \sigma^2\}$ where $\eta_i=\beta_0+\bX_i\bbeta+\theta_i$ for $i=1,\dots,n$. We average the results over 50 simulations for each combination of $n$ and $\rho$.

\subsection{Results}
We report the results in Tables \ref{tab:n100rho01}-\ref{tab:n250rho5}. For estimating $\bbeta$, all methods perform comparably in terms of both MSE and coverage, with the R2D2 prior yielding slightly lower MSE values when $n=100$. The results appear to be independent of $\rho$, while MSE decreases as $n$ increases for all models. Additionally, the different R2D2 priors perform almost identically. The R2D2 priors perform notably better than the Vague and PC priors for estimating the spatial marginal variance $\sigma^2_\theta$. The Vague prior has large MSE values but good coverage. The PC prior performs better than the Vague prior when $n=100$, but slightly worse when $n=200$ in terms of MSE, and consistently has the lowest coverage. Again, the R2D2 priors are similar in terms of both MSE and coverage. Moreover for all methods, the MSE results do not decrease when $n$ increases to 200. As for estimating $\rho$, when $n=100$, the Vague and R2D2 priors yield the lowest MSEs. When $n=200$, however, the R2D2 prior slightly outperforms both Vague and PC. Each method yields slightly lower MSEs when $n$ is larger. Finally, MSE and coverage are comparable across methods for estimating $R^2_n$.

\subsection{Discussion}
We close the simulation study section with a brief discussion. First, each method yields comparable results for estimating $\bbeta$. Since each of the prior frameworks focuses on the spatial variance and range parameters, the similar results for estimating the fixed effects is to be expected. The main focus of the simulation study, however, is comparing the estimates of the spatial parameters. For $\sigma^2_\theta$, the Vague prior distribution has non-trivial mass at large values, which leads to overestimating $\sigma^2_\theta$ and thus high MSE. The PC and R2D2 priors, on the other hand, explicitly shrink the estimates of $\sigma^2_\theta$ which generally leads to lower MSE values. It is evident that the R2D2 prior has greater shrinkage as it has significantly lower MSE values than both Vague and PC. Across all setting, the Vague and R2D2 priors yield comparable MSE values and both outperform PC when estimating $\rho$. Since the PC prior forces the estimate of $\rho\to\infty$, this could be leading to a large, positive bias and thus, larger MSE. The R2D2 construction only indirectly affects the prior distribution of $\rho$ (since $W$'s prior distribution is conditional on $\rho$), so the similar estimation to the Vague prior is reasonable. Additionally, \cite{zhang2004inconsistent} proved that neither $\sigma^2_\theta$ nor $\rho$ are consistently estimable under the in-fill asymptotic setting. This explains why the MSE values for $\sigma^2_\theta$ do not decrease with increasing $n$ and while there is only a small decrease for $\rho$. Finally, each method yields similar estimates of $R^2_n$. Since this is largely driven by estimates of the fixed effects $\bbeta$ and response variance $\sigma^2$, and these are similar across models, this is a sensible result.

\begin{table}
  \centering
    \begin{tabular}{c|cccccccc}
    \multicolumn{1}{c|}{} & \multicolumn{2}{c}{$\bbeta$} & \multicolumn{2}{c}{$\sigma^2_\theta$} & \multicolumn{2}{c}{$\rho$}& \multicolumn{2}{c}{$R^2_n$} \\
     Method & MSE & Cov. & MSE & Cov. & MSE & Cov. & MSE & Cov. \\\hline
        Vague & 0.23 (0.01) & {\bf 0.96} & 0.30 (0.08) & 0.96 & 0.05 (0.01) & {\bf 1.00} & {\bf 0.06} (0.01) & {\bf 0.98} \\
        PC & 0.23 (0.01) & 0.95 & 0.25 (0.01) & 0.34 & 0.08 (0.00) & 0.74 & 0.06 (0.01) & 0.80  \\
        R2D2 (1,1) & 0.21 (0.01) & 0.94 & {\bf 0.13} (0.01) & {\bf 1.00} & {\bf 0.04} (0.01) & {\bf 1.00} & 0.07 (0.01) & 0.94   \\
        R2D2 (1,4) & 0.22 (0.01) & 0.93 & 0.14 (0.01) & 0.92 & 0.05 (0.01) & {\bf 1.00} & 0.09 (0.01) & 0.86   \\
        R2D2 (4,1) & {\bf 0.20} (0.01) & 0.95 & 0.14 (0.02) & {\bf 1.00} & 0.05 (0.01) & {\bf 1.00} & 0.06 (0.01) & 0.96
    \end{tabular}
    \caption{Simulation study results for $n=100$ and $\rho=0.1$. MSE is root mean squared error loss with posterior median and Cov is frequentist coverage of 95\% posterior credible interval. Standard errors are in parentheses. The best values for each column are in {\bf bold}.}
    \label{tab:n100rho01}
\end{table}

\begin{table}
  \centering
    \begin{tabular}{c|cccccccc}
    \multicolumn{1}{c|}{} & \multicolumn{2}{c}{$\bbeta$} & \multicolumn{2}{c}{$\sigma^2_\theta$} & \multicolumn{2}{c}{$\rho$}& \multicolumn{2}{c}{$R^2_n$} \\
     Method & MSE & Cov. & MSE & Cov. & MSE & Cov. & MSE & Cov. \\\hline
        Vague & 0.22 (0.01) & {\bf 0.96} & 0.30 (0.05) & {\bf 1.00} & 0.09 (0.01) & {\bf 1.00} & 0.06 (0.01) & 0.94  \\
        PC  & 0.23 (0.01) & 0.94 & 0.25 (0.00) & 0.32 & 0.16 (0.00) & 0.68 & 0.05 (0.01) & 0.88   \\
        R2D2 (1,1) & 0.21 (0.01) & 0.95 & {\bf 0.10} (0.01) & {\bf 1.00} & 0.07 (0.01) & {\bf 1.00} & 0.04 (0.01) & 0.98  \\
        R2D2 (1,4) & 0.21 (0.01) & 0.94 & 0.10 (0.01) & {\bf 1.00} & 0.07 (0.01) & {\bf 1.00} & 0.05 (0.01) & 0.98    \\
        R2D2 (4,1) & {\bf 0.20} (0.01) & 0.95 & 0.12 (0.01) & {\bf 1.00} & 0.08 (0.01) & {\bf 1.00} & {\bf 0.04} (0.00) & {\bf 1.00}
    \end{tabular}
    \caption{Simulation study results for $n=100$ and $\rho=0.2$. MSE is root mean squared error loss with posterior median and Cov is frequentist coverage of 95\% posterior credible interval. Standard errors are in parentheses. The best values for each column are in {\bf bold}.}
    \label{tab:n100rho02}
\end{table}

\begin{table}
  \centering
    \begin{tabular}{c|cccccccc}
    \multicolumn{1}{c|}{} & \multicolumn{2}{c}{$\bbeta$} & \multicolumn{2}{c}{$\sigma^2_\theta$} & \multicolumn{2}{c}{$\rho$}& \multicolumn{2}{c}{$R^2_n$} \\
     Method & MSE & Cov. & MSE & Cov. & MSE & Cov. & MSE & Cov. \\\hline
        Vague & 0.23 (0.01) & {\bf 0.96} & 0.25 (0.04) & 0.98 & 0.14 (0.01) & {\bf 1.00} & 0.05 (0.01) & 0.94\\
        PC & 0.24 (0.01) & 0.95 & 0.23 (0.01) & 0.44 & 0.24 (0.00) & 0.58 & 0.04 (0.01) & 0.86 \\
        R2D2 (1,1) & 0.21 (0.01) & 0.94 & 0.11 (0.01) & 0.98 & 0.14 (0.01) & {\bf 1.00} & 0.04 (0.01) & 0.92 \\
        R2D2 (1,4) & 0.21 (0.01) & 0.93 & {\bf 0.11} (0.01) & 0.96 & {\bf 0.13} (0.01) & {\bf 1.00} & 0.05 (0.01) & 0.94  \\
        R2D2 (4,1) & {\bf 0.21} (0.01) & 0.94 & 0.13 (0.02) & {\bf 1.00} & 0.14 (0.01) & {\bf 1.00} & {\bf 0.04} (0.01) & {\bf 0.96}
    \end{tabular}
    \caption{Simulation study results for $n=100$ and $\rho=0.3$. MSE is root mean squared error loss with posterior median and Cov is frequentist coverage of 95\% posterior credible interval. Standard errors are in parentheses. The best values for each column are in {\bf bold}.}
    \label{tab:n100rho5}
\end{table}

\begin{table}
  \centering
    \begin{tabular}{c|cccccccc}
    \multicolumn{1}{c|}{} & \multicolumn{2}{c}{$\bbeta$} & \multicolumn{2}{c}{$\sigma^2_\theta$} & \multicolumn{2}{c}{$\rho$}& \multicolumn{2}{c}{$R^2_n$} \\
     Method & MSE & Cov. & MSE & Cov. & MSE & Cov. & MSE & Cov. \\\hline
        Vague & 0.16 (0.01) & 0.94 & 0.23 (0.07) & {\bf 0.98} & 0.06 (0.01) & {\bf 1.00} & 0.04 (0.01) & 0.94  \\
        PC & 0.16 (0.01) & 0.94 & 0.29 (0.05) & 0.70 & 0.07 (0.00) & 0.94 & 0.06 (0.01) & 0.80  \\
        R2D2 (1,1) & 0.15 (0.00) & 0.94 & 0.13 (0.02) & 0.94 & 0.05 (0.01) & {\bf 1.00} & 0.04 (0.01) & 0.94  \\
        R2D2 (1,4) & 0.15 (0.00) & {\bf 0.95} & {\bf 0.12} (0.01) & 0.92 & 0.05 (0.01) & {\bf 1.00} & 0.04 (0.00) & {\bf 0.96}  \\
        R2D2 (4,1) &  {\bf 0.15} (0.00) & 0.94 & 0.14 (0.03) & {\bf 0.98} & {\bf 0.05} (0.01) & {\bf 1.00} & {\bf 0.04} (0.01) & {\bf 0.96}
    \end{tabular}
    \caption{Simulation study results for $n=200$ and $\rho=0.1$. MSE is root mean squared error loss with posterior median and Cov is frequentist coverage of 95\% posterior credible interval. Standard errors are in parentheses. The best values for each column are in {\bf bold}.}
    \label{tab:n250rho1}
\end{table}

\begin{table}
  \centering
    \begin{tabular}{c|cccccccc}
    \multicolumn{1}{c|}{} & \multicolumn{2}{c}{$\bbeta$} & \multicolumn{2}{c}{$\sigma^2_\theta$} & \multicolumn{2}{c}{$\rho$}& \multicolumn{2}{c}{$R^2_n$} \\
     Method & MSE & Cov. & MSE & Cov. & MSE & Cov. & MSE & Cov. \\\hline
        Vague & 0.16 (0.01) & 0.93 & 0.19 (0.03) & 0.94 & 0.11 (0.01) & {\bf 1.00} & 0.03 (0.00) & 0.96 \\
        PC & 0.16 (0.01) & {\bf 0.94} & 0.21 (0.02) & 0.84 & 0.12 (0.01) & 0.96 & 0.03 (0.00) & {\bf 0.98}\\
        R2D2 (1,1) &0.15 (0.01) & 0.93 & 0.11 (0.01) & {\bf 1.00} & 0.07 (0.01) & {\bf 1.00} & 0.02 (0.00) & {\bf 0.98}\\
        R2D2 (1,4) & 0.16 (0.01) & 0.92 & 0.11 (0.01) & {\bf 1.00} & {\bf 0.07} (0.01) & {\bf 1.00} & 0.03 (0.00) & 0.96 \\
        R2D2 (4,1) & {\bf 0.15} (0.01) & 0.93 & {\bf 0.11} (0.01) & {\bf 1.00} & 0.07 (0.01) & {\bf 1.00} & {\bf 0.02} (0.00) & {\bf 0.98}
    \end{tabular}
    \caption{Simulation study results for $n=200$ and $\rho=0.2$. MSE is root mean squared error loss with posterior median and Cov is frequentist coverage of 95\% posterior credible interval. Standard errors are in parentheses. The best values for each column are in {\bf bold}.}
    \label{tab:n250rho2}
\end{table}

\begin{table}
  \centering
    \begin{tabular}{c|cccccccc}
    \multicolumn{1}{c|}{} & \multicolumn{2}{c}{$\bbeta$} & \multicolumn{2}{c}{$\sigma^2_\theta$} & \multicolumn{2}{c}{$\rho$}& \multicolumn{2}{c}{$R^2_n$} \\
     Method & MSE & Cov. & MSE & Cov. & MSE & Cov. & MSE & Cov. \\\hline
        Vague & 0.15 (0.01) & {\bf 0.97} & 0.15 (0.02) & 0.98 & 0.12 (0.01) & {\bf 1.00} & 0.04 (0.01) & 0.90\\
        PC & 0.15 (0.01) & 0.96 & 0.19 (0.01) & 0.90 & 0.19 (0.01) & 0.88 & 0.04 (0.00) & 0.94 \\
        R2D2 (1,1) & 0.14 (0.01) & 0.96 & 0.11 (0.01) & 0.98 & 0.11 (0.01) & {\bf 1.00} & 0.03 (0.00) & 0.94 \\
        R2D2 (1,4) & 0.14 (0.01) & 0.95 & {\bf 0.10} (0.01) & 0.96 & {\bf 0.10} (0.01) & {\bf 1.00} & 0.04 (0.00) & {\bf 0.98}\\
        R2D2 (4,1) & {\bf 0.14} (0.01) & 0.96 & 0.11 (0.01) & {\bf 1.00} & 0.10 (0.01) & 0.98 & {\bf 0.03} (0.00) & 0.94 
    \end{tabular}
    \caption{Simulation study results for $n=200$ and $\rho=0.3$. MSE is root mean squared error loss with posterior median and Cov is frequentist coverage of 95\% posterior credible interval. Standard errors are in parentheses. The best values for each column are in {\bf bold}.}
    \label{tab:n250rho5}
\end{table}

\clearpage

\section{MPA data analysis}

\subsection{Sensitivity analysis}
We re-run the real-data analysis for some other hyper-parameter combinations to show that the results are insensitive to these choices. We used 110,000 MCMC samples with the first 10,000 as burn-in and thinned the chain by keeping every fifth sample. The results are in Table \ref{tab:mpa}. There is small change in parameter estimates between the different settings despite large differences in the prior distributions as the prior for $\rho$ is on the log-scale.

\begin{table}[]
    \centering
 \begin{tabular}{c|cccc}
        \hline
         Setting & $R^2_n$ & $\beta_1$ & $W$ & $\sigma^2_\theta$ \\\hline
         1 &  0.42 (0.35, 0.50) & 0.07 (-0.05, 0.20) & 0.92 (0.48, 1.88) & 0.56 (0.26, 1.30)\\
         2 & 0.43 (0.35, 0.51) & 0.07 (-0.05, 0.20) & 0.92 (0.48, 1.96) & 0.52 (0.20, 1.30)\\
         3 & 0.45 (0.36, 0.53) & 0.07 (-0.05, 0.20) & 0.91 (0.50, 1.71) & 0.50 (0.22, 1.04) \\\hline
        Setting & $\rho$ & $\phi_1$ & $\phi_2$ & $\phi_3$ \\\hline
        1 &  0.07 (0.01, 0.36) & 0.17 (0.04, 0.44) & 0.17 (0.04, 0.42) & 0.63 (0.36, 0.86)\\
        2 & 0.05 (0.01, 0.55) & 0.19 (0.04, 0.51) & 0.18 (0.04, 0.50) & 0.60 (0.26, 0.85)\\ 
        3 & 0.02 (0.00, 0.15) & 0.20 (0.05, 0.51) & 0.19 (0.03, 0.51) & 0.57 (0.25, 0.84)
    \end{tabular}
    \caption{Posterior median and 95\% credible intervals for R2D2 prior sensitivity analysis with $R^2_n\sim\mathsf{Beta}(1,1)$ for the MPA fisheries data. Setting 1 (original): $a_0=b_0=0.10$, $\log\rho\sim\mathsf{Normal}(-2,1)$; Setting 2: $a_0=b_0=0.01, \log\rho\sim\mathsf{Normal}(-1,4)$; Setting 3: $a_0=b_0=0.01, \log\rho\sim\mathsf{Normal}(-3,4)$. All other hyper-parameters were as in the original prior formulation.}
    \label{tab:mpa}
\end{table}

\subsection{Trace plots for MPA data}
\begin{figure}
\centering
\begin{subfigure}{.45\textwidth}
  \centering
  \includegraphics[width=.99\linewidth]{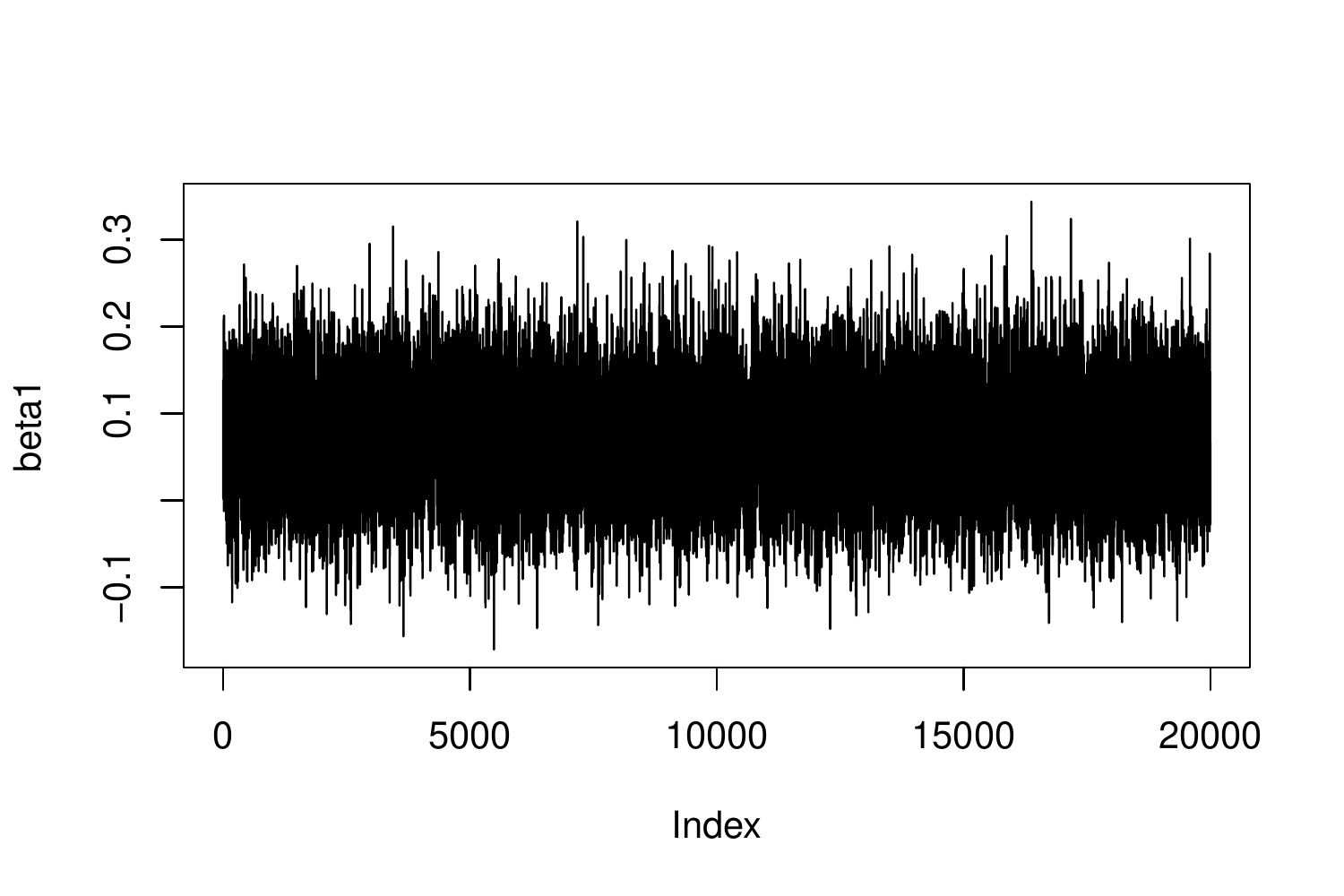}
  \caption{$\beta_1$}
  \label{fig:sub1}
\end{subfigure}%
\begin{subfigure}{.45\textwidth}
  \centering
  \includegraphics[width=.99\linewidth]{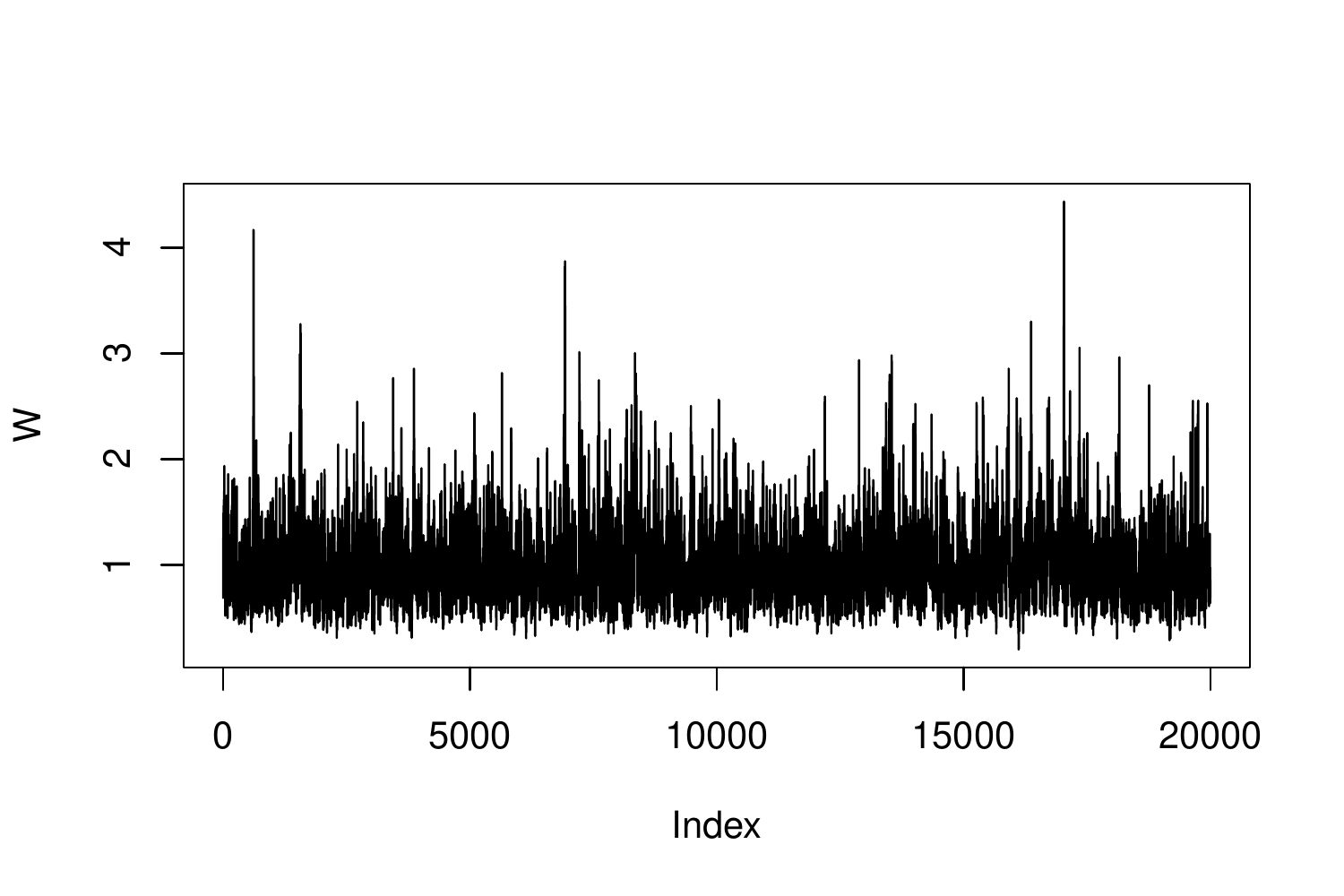}
  \caption{$W$}
  \label{fig:sub2}
\end{subfigure}\\
\begin{subfigure}{.45\textwidth}
  \centering
  \includegraphics[width=.99\linewidth]{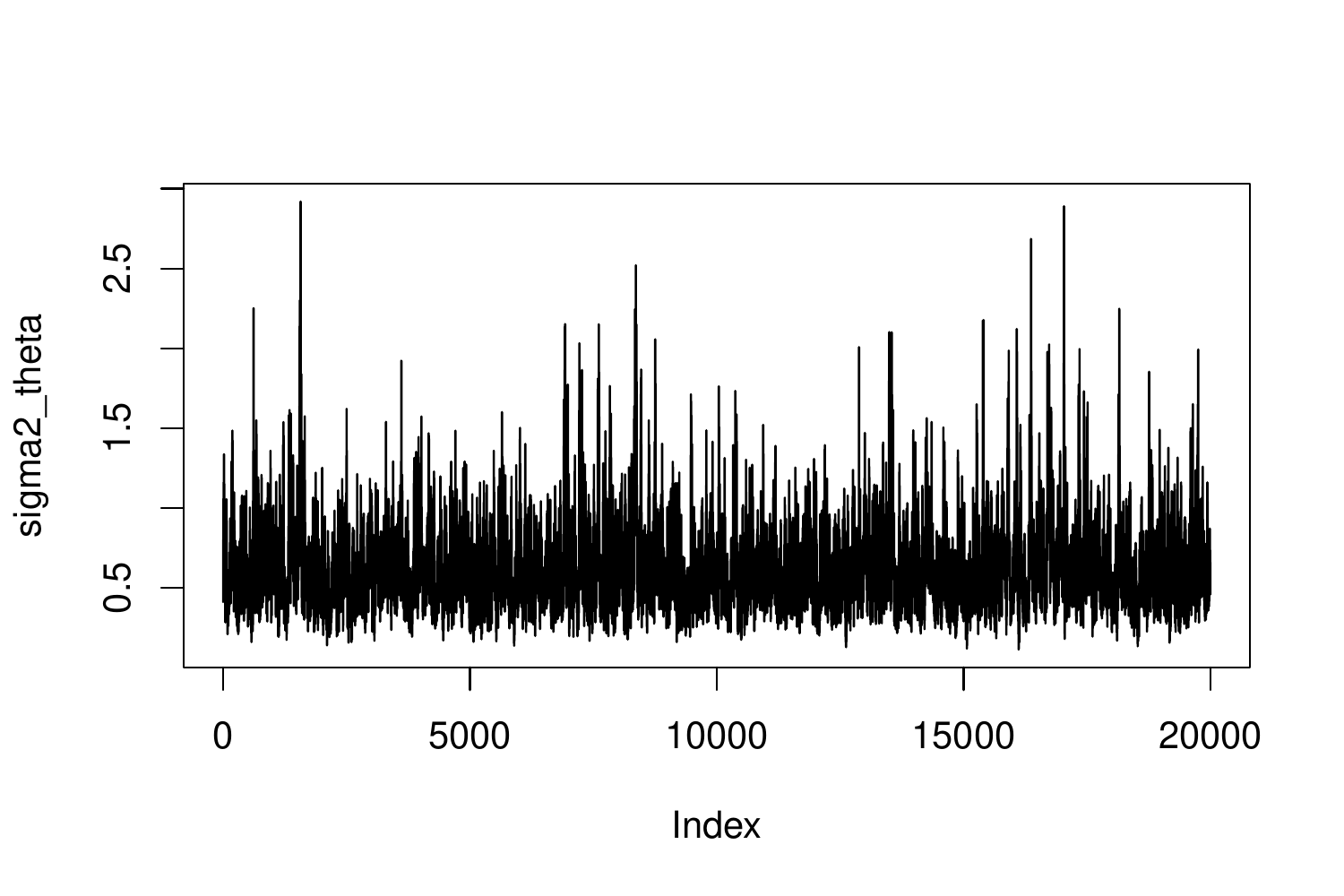}
  \caption{$\sigma^2_\theta$}
  \label{fig:sub1}
\end{subfigure}%
\begin{subfigure}{.45\textwidth}
  \centering
  \includegraphics[width=.99\linewidth]{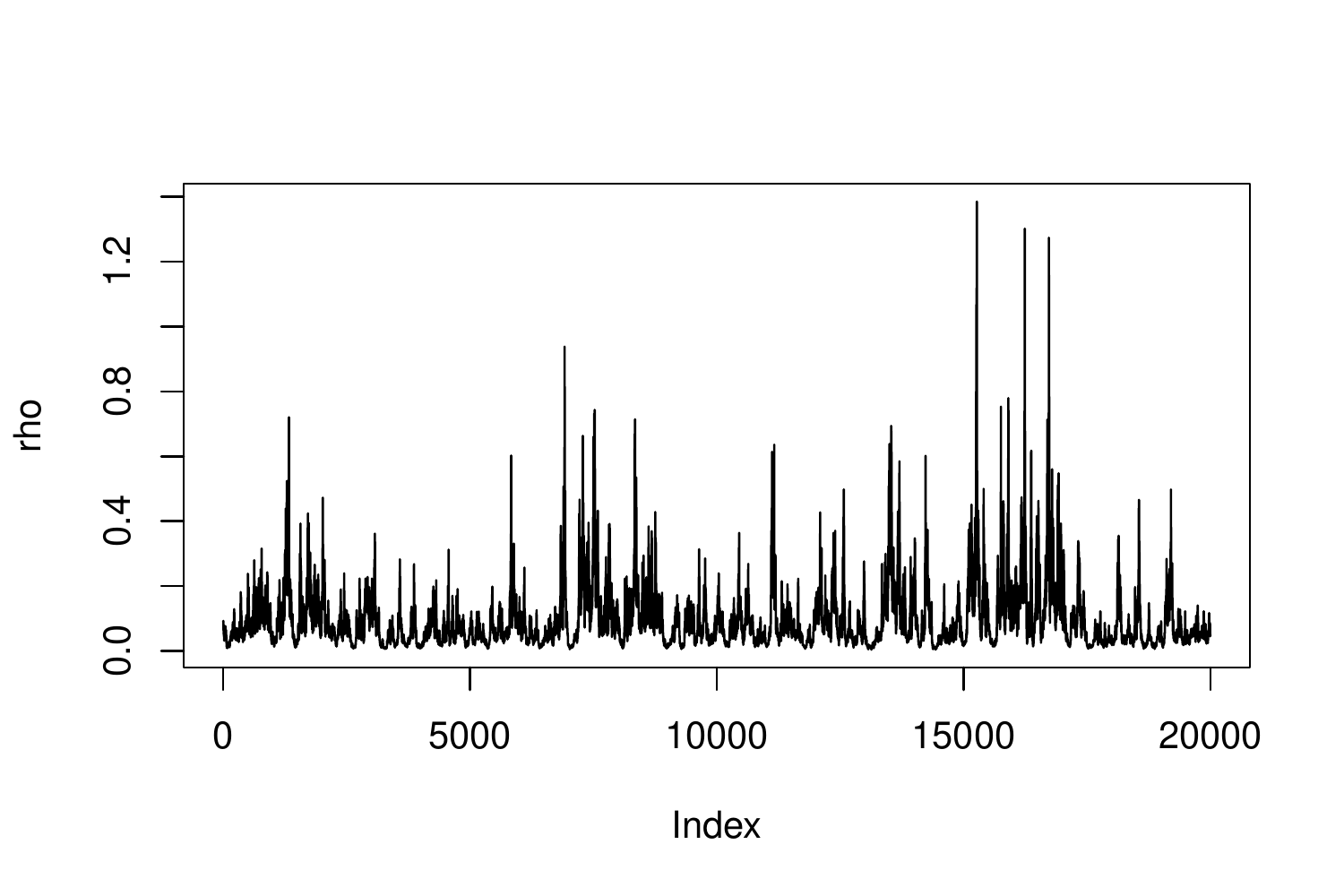}
  \caption{$\rho$}
  \label{fig:sub2}
\end{subfigure}\\
\begin{subfigure}{.45\textwidth}
  \centering
  \includegraphics[width=.99\linewidth]{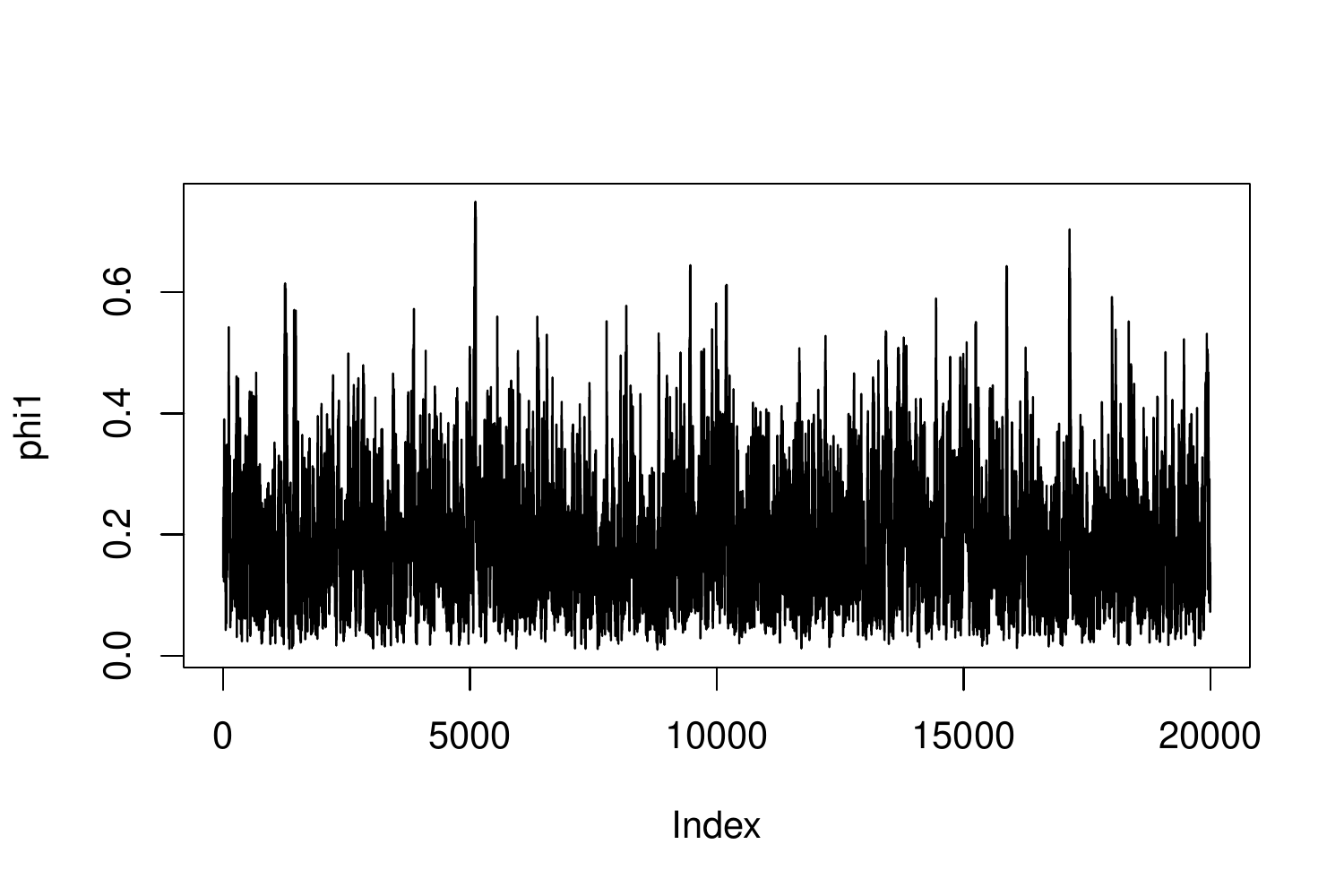}
  \caption{$\phi_1$}
  \label{fig:sub2}
\end{subfigure}
\caption{Trace plots for various parameters of the MPA data set using the $R^2_n\sim\mathsf{Beta}(1,1)$ prior.}
\label{fig:trace}
\end{figure}

See Figure \ref{fig:trace} for trace plots of various parameters of the MPA data set using the $R^2_n\sim\mathsf{Beta}(1,1)$ prior. Other priors led to similar convergence.

\clearpage

\subsection{Effective samples}
We report the number of effective samples in the MCMC chain for several parameters and each method in Table \ref{tab:ess}.

\begin{table}[]
    \centering
    \begin{tabular}{c|ccccc}
         $(a,b)$ & time  & $\beta_1$ & $W$ & $\sigma^2_\theta$ & $\rho$   \\\hline
         Vague & 270 & 20,000  & $-$ & 4500 & 210 \\ 
         PC & 350 & 18,000 & $-$ & 4400  & 200  \\
         $(1,1)$ & 490 & 14,800 & 1500 & 1000 & 240 \\
         $(\tfrac12,\tfrac12)$ &480 & 13,600 &1200 &1000 &250\\
         $(1,4)$ & 490 & 15,300 & 1700 &1000 & 290\\
         $(4,1)$ & 490 & 14,100 & 1400 &900 & 240
    \end{tabular}
    \caption{Run time (in minutes) and number of effective samples for the MCMC chain for several model parameters. Note that the run time includes burn-in samples.}
    \label{tab:ess}
\end{table}

\subsection{Fixed effect estimates}
We also report the posterior mean and 95\% credible intervals for the each of the $p=9$ fixed effects for the MPA data analysis in Tables \ref{tab:fixed1}, \ref{tab:fixed2} and \ref{tab:fixed3}. In general, the Vague and PC priors result in fixed effects with greater magnitude. In particular, consider the estimate for Sea Temperature where the Vague and PC prior's estimates are larger than those of R2D2. This is because the R2D2 priors assumed that the variance was the {\it same} for all fixed effects, i.e., $\beta_j\sim \mathsf{Normal}(0, \sigma^2\phi_1W/p)$ for all $j$. Since most fixed effects estimates are small in magnitude, the variance is small which means it is difficult for the any one effect estimate to be large. So while the estimate of Sea Temperature is the largest of all the fixed effects, it will not be as large as the estimate from the Vague prior. If each fixed effect was allowed to have its own variance, i.e., $\beta_j\sim \mathsf{Normal}(0,\sigma^2\phi_jW)$, then it is likely that the estimate for Sea Temperature would be larger.

\begin{table}[h]
    \centering
    \begin{tabular}{l|cc|cc|cc}
    \multicolumn{1}{l|}{} &
    \multicolumn{2}{c|}{MU (0) vs. NT (1)}& 
    \multicolumn{2}{c|}{Depth}& 
    \multicolumn{2}{c}{Wave exposure} \\
    Prior & Mean & 95\% CI  & Mean &95\% CI & Mean &95\% CI \\\hline
    Vague & 0.11 & (-0.03, 0.24) & 0.18 & (0.05, 0.31) & 0.04 & (-0.10, 0.17)  \\
    PC & 0.10 & (-0.03, 0.24) & 0.19 & (0.06, 0.32) & 0.04 & (-0.10, 0.18) \\
    $R^2_n\sim\mathsf{Beta}(1,1)$ & 0.07 & (-0.05, 0.20) & 0.17 & (0.04, 0.29) & 0.04 & (-0.09, 0.16)  \\
    $R^2_n\sim\mathsf{Beta}(0.5,0.5)$ & 0.07 & (-0.05, 0.2) & 0.17 & (0.05, 0.29) & 0.04 & (-0.09, 0.16) \\
    $R^2_n\sim\mathsf{Beta}(1,4)$ & 0.07 & (-0.05, 0.19) & 0.17 & (0.05, 0.29) & 0.04 & (-0.08, 0.16) \\
    $R^2_n\sim\mathsf{Beta}(4,1)$ & 0.07 & (-0.05, 0.20) & 0.17 & (0.05, 0.30) & 0.04 & (-0.09, 0.17) 
    \end{tabular}
        \caption{Posterior median and 95\% credible intervals for fixed effects in MPA data analysis.}
    \label{tab:fixed1}
\end{table}
\begin{table}[h]
    \centering
    \begin{tabular}{l|cc|cc|cc}
    \multicolumn{1}{l|}{} &
    \multicolumn{2}{c|}{Dist. to shoreline}& 
    \multicolumn{2}{c|}{Dist. to market}& 
    \multicolumn{2}{c}{Population}\\
    Prior & Mean & 95\% CI  & Mean &95\% CI & Mean &95\% CI  \\\hline
    Vague & 0.03 & (-0.18, 0.25) & 0.11 & (-0.55, 0.86) & 0.00 & (-0.34, 0.36) \\
    PC & 0.03 & (-0.17, 0.24) & 0.11 & (-0.51, 0.78) & 0.00 & (-0.33, 0.33) \\
    $R^2_n\sim\mathsf{Beta}(1,1)$ & 0.03 & (-0.14, 0.21) & 0.06 & (-0.19, 0.34) & -0.01 & (-0.23, 0.22) \\
    $R^2_n\sim\mathsf{Beta}(0.5,0.5)$ & 0.04 & (-0.14, 0.21) & 0.06 & (-0.19, 0.35) & -0.01 & (-0.23, 0.22)\\
    $R^2_n\sim\mathsf{Beta}(1,4)$ & 0.03 & (-0.13, 0.20) & 0.06 & (-0.18, 0.33) & -0.01 & (-0.21, 0.20)  \\
    $R^2_n\sim\mathsf{Beta}(4,1)$ & 0.04 & (-0.14, 0.22) & 0.07 & (-0.20, 0.38) & -0.01 & (-0.24, 0.23) 
    \end{tabular}
    \caption{Posterior median and 95\% credible intervals for fixed effects in MPA data analysis.}
    \label{tab:fixed2}
\end{table}
\begin{table}[h]
    \centering
    \begin{tabular}{l|cc|cc|cc}
    \multicolumn{1}{l|}{} &  
    \multicolumn{2}{c|}{Sea temp}& 
    \multicolumn{2}{c|}{Chlorophyll-a}& 
    \multicolumn{2}{c}{Reef area} \\
    Prior & Mean & 95\% CI  & Mean &95\% CI & Mean &95\% CI \\\hline
    Vague & 1.07 & (0.32, 2.17) & 0.09 & (-0.12, 0.32) & 0.09 & (-0.14, 0.30)   \\
    PC & 0.94 & (0.29, 1.94) & 0.08 & (-0.14, 0.29) & 0.08 & (-0.13, 0.30)   \\
    $R^2_n\sim\mathsf{Beta}(1,1)$ & 0.21 & (-0.04, 0.60) & -0.01 & (-0.17, 0.17) & 0.07 & (-0.11, 0.25)  \\
    $R^2_n\sim\mathsf{Beta}(0.5,0.5)$ &  0.22 & (-0.04, 0.64) & 0.00 & (-0.17, 0.17) & 0.07 & (-0.11, 0.25)\\
    $R^2_n\sim\mathsf{Beta}(1,4)$ & 0.20 & (-0.04, 0.57) & -0.01 & (-0.17, 0.16) & 0.07 & (-0.10, 0.24) \\
    $R^2_n\sim\mathsf{Beta}(4,1)$ & 0.24 & (-0.04, 0.67) & 0.00 & (-0.18, 0.18) & 0.07 & (-0.11, 0.26)  
    \end{tabular}
    \caption{Posterior median and 95\% credible intervals for fixed effects in MPA data analysis.}
    \label{tab:fixed3}
\end{table}




%


\end{document}